\documentclass{article}
\usepackage[margin=1in]{geometry}
\usepackage[utf8]{inputenc}
\usepackage{amsmath}
\usepackage[numbers]{natbib}
\usepackage{hyperref}
\usepackage{url}
\usepackage{graphicx}
\usepackage{booktabs}
\usepackage{lipsum}
\usepackage{float}
\usepackage[usenames,dvipsnames]{xcolor}
\usepackage[linesnumbered,ruled,vlined]{algorithm2e}
\usepackage{siunitx}
\usepackage{subcaption}
\usepackage{alltt}

\title{An Comparative Analysis of Different Pitch and Metrical Grid Encoding Methods in the Task of Sequential Music Generation}
\author{
Yuqiang Li\thanks{Xi'an Jiaotong-Liverpool Universiy} \quad
Shengchen Li\thanks{Xi'an Jiaotong-Liverpool Universiy} \quad
George Fazekas\thanks{Queen Mary University of London}
\\
\small \texttt{\{yuqiang.li19@student., shengchen.li@\}xjtlu.edu.cn}\\\small \texttt{g.fazekas@qmul.ac.uk}
}

\date{}

\begin{document}

\maketitle

\section*{Abstract}
Pitch and meter are two fundamental music features for symbolic music generation tasks,
where researchers usually choose different encoding methods depending on specific goals.
However, the advantages and drawbacks of different encoding methods have not been frequently discussed.
This paper presents a integrated analysis of the influence of two low-level feature, pitch and meter, on the performance of a token-based sequential music generation model.
First, the commonly used MIDI number encoding and a less used pitch class-octave encoding are compared.
Second, an dense intra-bar metric grid is imposed to the encoded sequence as auxiliary features.
Different complexity and resolution settings of the metric grid are compared.
For complexity, the single token approach and the multiple token approach are compared;
for grid resolution, 0 (ablation), 1 (bar-level), 4 (downbeat-level) 12, (8th-triplet-level) up to 64 (64th-note-grid-level) are compared;
for durational resolution, 4, 8, 12 and 16 subdivisions per beat are compared.
All different encodings are tested on separately trained Transformer-XL model for a melody generation task.
From the perspective of distribution distance of several objective evaluative metrics to the test dataset,
the results suggest that the class-octave encoding significantly outperforms the taken-for-granted MIDI encoding on pitch-related metrics;
higher grid resolutions and multiple-token grid also significantly increases the generation quality of the rhythm,
but also suffer from over-fitting at the early stage of training.
Results also display a general phenomenon of over-fitting from two aspects, the pitch embedding space and the test loss of the single-token grid encoding.
From a practical perspective, we both demonstrate the feasibility and raise the concern of easy over-fitting problem of using smaller networks and lower embedding dimensions on the generation task,
The finding can also contribute to futural models in terms of feature engineering.

\paragraph{Keywords}  Music Representation, Music Generation, Pitch, Rhythm, Feature Engineering

\section{Introduction}\label{sec:intro}

Symbolic music representation is to some extent in between the standard music notation system and the actual sound of performed music. In the computer audition tasks, it has the advantages of abstraction and conciseness compared to the detailed wave form as used in the acoustic domain.
The abstraction in symbolic representations allows the annotations of higher-level musical features, including form, expressiveness, and articulations, besides the fundamental pitch, harmonic, metric and rhythmic elements~\citep{wigginsFrameworkEvaluationMusic1993}.

Specifically when used in machine learning, the sparsity and conciseness of symbolic representation are further required depending on the model's representation capacity and their specific limitation on the computational resources.
Hence, instead of directly using score-level representations, such as MusicXML and ABC notation, researchers proposed a wide range of methods to select minimal features and encode them in a new representation that enables a specific model to learn and generalize well.
Low-level features, including pitch, duration and velocity can be effortlessly obtained from both MIDI event sequence and score notations. These features are commonly encoded into matrices~\citep{dongMuseGANMultitrackSequential2017}, word sequences~\citep{huangPopMusicTransformer2020}, vectors with certain geometric constraints~\cite{chewMathematicalModelTonality2000}, graphs~\citep{jeongGraphNeuralNetwork2019,hsiaoCompoundWordTransformer2021} or different forms before being processed by a model.

Despite the musical information carried by low-level features, it is still challenging for the latest symbolic music algorithmic composition models to directly learn from such features and generate music to a satisfying extent.
For instance, most models have been struggling with modeling long-term temporal dependency of music~\citep{herremansFunctionalTaxonomyMusic2018}, although this problem could be alleviated by providing the model with explicit structural information~\citep{wuJazzTransformerFront2020,daiCONTROLLABLEDEEPMELODY2021,luMeloFormGeneratingMelody2022}.
Also, the generation systems suffer from the lack of semantic representation which can be interpreted musicologically.

Recent works have displayed a trend of introducing more prior knowledge to a machine learning model by applying some basic music theories.
From a feature engineering perspective, we categorize these works into three types: feature aggregation, feature selection and feature encoding.
Feature aggregation seems to be the most investigated respect in recent works, by which we refer to practices that manipulate the selected features inside the model so as to add constraints or create topological connections according to music theories~\citep{huangMusicTransformer2019,huangPopMusicTransformer2020,wangPIANOTREEVAEStructured2020,lieckTonalDiffusionModel2020,hsiaoCompoundWordTransformer2021,songSINTRALEARNINGINSPIRATION2021}.
As to feature selections, there are works utilizing higher-level features from either mathematical calculation or extra labels annotated by experts or musicians~\citep{wuJazzTransformerFront2020,zengMusicBERTSymbolicMusic2021,luMeloFormGeneratingMelody2022}.
Feature encoding focuses on how features are converted to numerical values~\citep{liangPiRhDyLearningPitch2020,lazzariPitchclass2vecSymbolicMusic2022,guoDomainKnowledgeInspiredMusicEmbedding2022}, but it is less mentioned in the current literature compared to the other two respects.

Therefore, our work focuses on feature encoding and attempts to systematically compare several existing encoding methods of the low-level pitch and metric (positional) features.
This study is based on the monophonic melody generation task, using only low-level features as model input, and particularly, only
flat event-like sequential representations, out of the following concerns.
First, lower-level features are more independent and thus more controllable.
It is relatively easier to isolate certain features and investigate the importance of different features.
Second, melodies are monophonic, which are much simpler to model compared to polyphonic music where both harmony and inter-part interactions must be considered.
Third, the expressiveness (e.g., change in the tempo and dynamics) in the music is ignored since they bring extra temporal dimensions that compound and interfere with the low-level features, which is beyond the scope of this study.
Finally, preferring flat sequential representation over any advanced topological representations (e.g., stacking tokens into super-tokens as in~\citep{hsiaoCompoundWordTransformer2021,zengMusicBERTSymbolicMusic2021} or using hierarchical representations as in~\citep{wangPIANOTREEVAEStructured2020} )
helps avoid introducing new feature selection bias and new hyper-parameters.
An extra gain of such concern is that other futural models can easily build up on the results of this study,
since only some minor modifications on the input data representation could lead to noticeable improvement on the model performance.

Regarding pitch, duration and metric features, this study discusses four hyper-parameters involved in their encodings.
(1) \textit{Pitch} encoding, including the commonly used MIDI pitch number, and the pitch class-pitch octave pair as used by~\cite{liangPiRhDyLearningPitch2020}.
(2) (Bar-level grid) \textit{Position Complexity} (PC), referring to whether a single token or multiple tokens are used to represent different metric grid positions inside a bar.
(3) (Bar-level grid) \textit{Position Resolution} (PR), the number of evenly distributed positions of a bar to be encoded.
(4) (Note) \textit{Duration Resolution} (DR), the number of subdivisions of a beat, which determines the minimal unit of note length.
It is hypothesized that these 4 hyper-parameters greatly influence the model performance.
It is expected that more complicated token representations, that is, pitch class-octave encoding, multiple-token PC, higher PR and DR would result in better generation quality in terms of how well they approximate true distribution of several selected objective evaluation metrics.

In order to test the hypothesis, we first define a few possible options for our four hyper-parameters.
A brute force searching is then conducted on the hyper-parameter grid, that is, all the possible configurations of the hyper-parameter grid are individually used to transform the Wikifonia dataset and train the same Transformer-XL melody generation model from scratch for the same number of gradient updates.
Objective analysis of the generation quality is done by first sampling a large number of melodies from the resulted models,
then comparing their metric distribution similarity by the Overlapped Area (OA) with the test set distribution,
in terms of 9 evaluative scoring metrics: 5 pitch-related and 4-rhythmic features.

Regarding pitch encoding, the paired \(t\)-test results report a significantly better average performance of the class-octave encoding over the commonly used MIDI number encoding.
For PC, PR and DR, they interact in a way that higher PR and DR combined with single-token PC resulted in the best approximations.

The interaction of different encoding hyper-parameters are discussed in Section~\ref{sec:combined} followed by a discussion on the over-fitted pitch embedding space.

The main contributions of this work are twofold.
First, we demonstrate that a small Transformer-XL network of only 0.5M parameters and a low dimensional (\(d=32\)) embedding space are able to produce music with close objective metric distributions, given appropriate encoding hyper-parameters and a few epochs of training.
The advantages and drawbacks of different encoding options are manifested by the results.
Second, we call attention to the over-fitting problem and the exposure bias due to the nature of the task, which also interact with the encoding hyper-parameters and influence the model performance.
We believe that the findings of this study could be easily extended to the improvement of other music generation models.

\section{Related Works}\label{sec:related}
\subsection{Pitch Feature Encodings}
Categorical pitch encoding seems to be the most used pitch encoding in the current symbolic music representations~\citep{briotDeepLearningTechniques2019},
which can be found in representations such as a pianoroll stacked by one-hot pitch vectors.
The problem of this encoding is that no explicit prior knowledge about pitches is encoded, since all the pitches are equidistant from the others.

Early attempts at addressing this issue were to construct a static pitch representation space that preserves the pitch similarities based on listener's ratings in the psychoacoustical experiments~\citep{krumhanslPsychologicalRepresentationMusical1979,krumhanslTracingDynamicChanges1982,shepardGeometricalApproximationsStructure1982}.
Based on these,~\cite{mozerCONNECTIONISTMUSICCOMPOSITION1990} created \textsl{concert}, a neural network-based music generation system with a proposed pitch representation named PHCCCF, out of three components of an absolute pitch: Pitch Height (PH), Chromatic Circle (CC), Circle of Fifths (CF).
~\cite{ladenRepresentationPitchNeural1989} compared a few pitch representations on a neural net chord classifier, including
the (categorical) pitch class representation and a few psychoacoustical pitch representations involving harmonics.
The results suggested that explicitly encoded pitch harmonics result in higher classification accuracy.

Recent solutions mainly favor the word embedding approach due to the rapid development of natural language processing (NLP) and the decent performance of the latest language models.
By word embedding, the vector representations of pitches are learned and can be dynamically optimized according to the downstream task.
This approach is commonly seen in MIDI event-based representations, such as the \textsc{Note-On}, \textsc{Note-Off} tokens in Performance RNN~\citeyear{performance-rnn-2017}, Music Transformer~\citep{huangMusicTransformer2019}; and \textsc{Note-On} in MusicVAE~\citep{robertsHierarchicalLatentVector2019}, REMI~\citep{huangPopMusicTransformer2020}, CWT~\citep{hsiaoCompoundWordTransformer2021} and MusicBERT~\citep{zengMusicBERTSymbolicMusic2021}.
However, the evaluation of pitch embedding space is rarely discussed in the literature. The choice of embedding dimension is usually empirically set to 512, which is taken for granted and we consider it unreasonably high.
\cite{guoDomainKnowledgeInspiredMusicEmbedding2022} proposes a low-level pitch embedding which ensure the translational invariance (or, transpositional invariance) that is not guaranteed by a trained word embedding.

Regarding pitch feature selection, relative pitch (interval, the delta pitch of two absolute pitches) could also be encoded~\citep{chewMathematicalModelTonality2000,lattnerLearningTranspositionInvariantInterval2018}.
However, when word embedding is used, the interaction between absolute pitch vectors and relative pitch vectors can unexplainable~\citep{guoDomainKnowledgeInspiredMusicEmbedding2022} and unpractical to use~\citep{huangMusicTransformer2019}.
Therefore, only absolute pitch is considered in our work.

The Tonnetz representation provide alternative geometric features of both pitch and interval, but it seems to appear more in non-generative tasks such as music classification~\citep{chewMathematicalModelTonality2000,chuanModelingTemporalTonal2018,lieckTonalDiffusionModel2020}.
As to pitch spelling,~\cite{micchiNotAllRoads2020} discussed the subtle differences between chromatic pitch (CP) and pitch spelling (PS) when encoding enharmonic notes (e.g., C\(\sharp\), D\(\flat\) and E\(\flat\flat\)) in the context of automatic harmonic analysis.

The pitch class feature seems to be seen more in discriminative tasks (music classification \citep{ladenRepresentationPitchNeural1989}, clustering \citep{yustClusteringBasedApproachAutomatic2022}, harmonic analysis \citep{micchiNotAllRoads2020}) rather than in generative models \citep{liangPiRhDyLearningPitch2020,lieckTonalDiffusionModel2020},
since most generative models to date still stick to the categorical encoding (e.g. 128 MIDI pitch numbers).
Hence, this study will compare the influence of these two different commonly used pitch encodings in the context of a generative model, which seems to be first work comparing them to the best of our knowledge. Specifically, we use a transformer-based sequential melody generation model.

\subsection{Duration Encodings}

\subsubsection{Implicit and Explicit Duration Encodings}
One factor of duration encoding is whether the note duration is encoded explicitly or implicitly.
An explicit encoding uses analogous numerical features for note length, which has been experimented since the very early model \textsc{concert} by  \cite{mozerCONNECTIONISTMUSICCOMPOSITION1990}.
Especially when analogous values are used for encodings, it allows different features to be calculated algebraically with an always meaningful duration interpretation.
When using word embedding to denote duration, it is usually discretized into finite different possible values,
as used in the recently proposed REMI representation \cite{huangPopMusicTransformer2020}.
An implicit encoding usually relies on a groups of tokens that accumulate the short time spans before the note is released.
This can be done using a single repetitive token that represents a fixed amount of time (more common in a non-expressive context), e.g., the \textsc{Hold} token as used in DeepBach \citep{hadjeresDeepBachSteerableModel2017},
or multiple tokens for different time spans, e.g., using the combination of different \textsc{Time-Shift} tokens and \textsc{Note-Off} tokens as in \citep{performance-rnn-2017}.

In REMI's work, \cite{huangPopMusicTransformer2020} concluded that explicit duration encoding outperformed the taken-for-granted implicit duration encoding (i.e., combination of \textsc{Time-Shift}s and \textsc{Note-Off}s), with better generation quality and shorter average sequence length.
However, in the comparison of the Baseline 1 and 2 where the only difference was implicit \textsc{Note-Off} versus explicit \textsc{Duration}, the resulted three objective evaluation metrics seem to be equally apart from the true distribution with one higher and the other lower, which might not result in a strong conclusion of the latter being better.
In our work, we will further compare this but using different terminology\footnote{Specifically, these two sets are referred by the settings of (Position Resolution = 0) and (Position Resolution = 4)} other than duration itself.

\subsubsection{Duration Resolution}
Another factor of duration encoding is the resolution.
The minimal step of time is usually defined by a hyperparameter, which we refer to as the \textit{resolution}, meaning the number of equal subdivisions of a beat or a bar and using one of them as the unit of time.
Presumably, researchers choose different \textit{beat resolution} because of the model capacity (e.g., the maximal sequence length of a model).
For example,
MuseGAN \citep{dongMuseGANMultitrackSequential2017} used 24 subdivisions of a beat on a deep convolutional generative adversarial network (DCGAN);
MidiNet \citep{yangMidiNetConvolutionalGenerative2017} used 4 on another DCGAN; MusicVAE \citep{robertsHierarchicalLatentVector2019} used 4 on a LSTM variational auto-encoder (VAE);
Pop Music Transformer \citep{huangPopMusicTransformer2020} reported the best performance of using 4 on a Transformer-XL.
Although increasing the beat resolution could allow more rhythmic details to be encoded, it would also potentially lead to longer sequence(especially for the implicit duration encodings),
whose advantages and drawbacks has not been extensively studied in the literature.
Therefore, this study will examine the concept of duration encoding from two aspects, namely \textit{duration (beat) resolution} (DR) and \textit{positional grid (beat) resolution} (PR) and investigate their influence on the model performance.
The definition of PR is given is given in the following subsection.

\subsection{Metrical Encodings and Bar-level Grid Position Encodings}
\subsubsection{Positional Resolution}
Since REMI \citep{huangPopMusicTransformer2020} and Jazz Transformer \citep{wuJazzTransformerFront2020},
it turns out effective to explicitly impose a bar-level metric grid\footnote{We will also use the term ``metric grid'' and ``positional grid'' interchangeably in the rest of this paper} on the encoded sequence to improve the generation quality and even increase the generation controllability.
According to these authors, most previous models were unable to generate pieces with clear pulses and beats.
Besides expressiveness features,
REMI proposed the grid-level metric encoding at both bar-level and the finest \textit{position}-level.
In REMI, \textit{position} refers to a series of special tokens (\textsc{pos}\(_{1..16}\)) indicating different possible grid positions inside each bar,
where the grid is evenly divided into 16 parts\footnote{By our terminology, the positional grid encoding above uses PR = 4, since 4 subdivisions are encoded for each beat.}.
Another special token, \textsc{bar}, refers to the first position of a bar (\textsc{Position}\(_1\)), but they use a different token that always comes before \textsc{pos}\(_1\) to emphasize the beginning of a bar.
The necessity of this \textsc{Bar} token deserves further discussion.
Technically, if all the absolute position tokens are strictly ordered in the encoded sequence, a single \textsc{Position (1/16)} is enough to indicate the begin of a new bar, in which case the \textsc{Bar} is optional.
In this study, we addressed these minor issues by ensuring the same amount of information being encoded inside each pair of candidates to compare.

Second, the comparison of Baseline 3 (a stronger baseline using explicit duration and multiple non-expressive \textsc{Time-Shifts}) and REMI (that uses multiple \textsc{Position}s and a single \textsc{Bar} token.) was designed for different encoding approaches but not based on the same amount of information.
Compared to Baseline 3, REMI provides extra information regarding bar lines and absolute positions, which means it is not feasible to reconstruct beat-level nor bar-level based on the encoded sequence for Baseline 3.

\subsubsection{Positional Complexity}
Similar to duration, the positional grid can also be implicit by accumulating the same token (e.g., Music Transformer generating Bach Chorale \citep{huangMusicTransformer2019}) or explicitly specified with absolute positions as used in REMI and the OctupleMIDI representation \citep{zengMusicBERTSymbolicMusic2021}.
In order to be distinguished from duration, we use the term \textit{Position Complexity} (PC) to indicate whether the grid is encoded by implicit accumulative single tokens or explicit multiple absolute positional tokens.
Correspondingly, we use \textit{single} and \textit{multiple} for the two options.

To summarize, this work attempts to decompose the encoding settings of low-level features (pitch, meters and )
The low-level presentation settings of some recent models are listed in table xxx.
Instead of vaguely using the concept of resolution, we try to decompose the encoding of duration and bar-level metric grid positions into 3 hyper-parameters,
PC, PR and DR, assuming that the note duration is encoded with explicit duration.

\section{Experiment Setup}

\subsection{Dataset and Preprocessing}
The Wikifonia dataset contains 6,405 lead sheets of music from mixed genres in the format of MusicXML.
A cleaned dataset is used, downloaded by the \texttt{muspy} library~\citep{dongMusPyToolkitSymbolic2020}.
As to time signature, only \(\frac{4}{4}\) is considered to avoid encoding inconsistency.
We removed songs containing inconsistent bar lengths (mostly because of the change of time signature).
90\% (3,861) samples were used for training and the other 10\% (429) samples were for the test set.
Chord, tempo, instrument and other metadata are all ignored in this study.

\subsection{Vocabulary}
The vocabulary set consists of three parts, pitch tokens (including \textsc{REST}, a special pitch token for silence), duration tokens and positional tokens,
whose specific tokens are defined by the four hyper-parameters.
The \textsc{Pitch} hyper-parameter determines pitch tokens with the \textsc{Number} and the \textsl{Class-Octave} option.
\textsc{Duration Resolution} (DR) determines the beat resolution which all the note onset and duration time are rounded to.
Duration tokens represent times from the smallest time step to 4 beats.
\textsc{Position Resolution} (PR) determines the amount of metric grid information is provided in the encoding and thus defines a set of positional tokens.
\textsc{Position Complexity} (PC) comes with 2 options, \textsc{Single} and \textsc{Multiple}, that specify whether to use a single token or multiple tokens for the different grid positions in each bar.
Other special tokens such as \textsc{PAD} would not be discussed in detail.

\subsection{Encoding Algorithm}
Given a melody \(M\) and an input representation vocabulary \(V\), we use the Algorithm~\ref{algo:encode} to encode a melody to a token sequence.

\begin{algorithm}
  \SetAlCapFnt{}
  \DontPrintSemicolon
  \KwIn{\(M\), a melody, \(V\), a vocabulary set}
  \KwOut{A list of encoded tokens}
  \textit{ticksPerStep} \(\gets\) \(M\).\textit{ticksPerQuarterNote} / \(V\).\textit{durationResolution}\;
  \(T \gets []\) \tcc*{A list of triple (time, tokenType, token)}
  Round all the onset time and duration of \(M\) to multiples of \textit{ticksPerStep}.\;
  Sort \(M\).notes by onset time in ascending order.\;
  \For{note \(\in M\).notes}{
    \(T\).extend(\(V\).encode\_pitch(\textit{note}))\;
    \(T\).extend(\(V\).encode\_duration(\textit{note}))\;
  }
  \If{\(V\).positionResolution \(>0\)}{
    \(T\).extend(\(V\).generate\_bar\_and\_grid\_tokens(\(M\).\textit{totalTime}))
  }
  \(T\).extend(\(V\).generate\_REST\_tokens\_to\_fill\_gaps(\(T\)))\;
  Sort \(T\) by\;
  \Indp primary key: \textit{time}, ascending\;
  secondary key: \textit{tokenType}, ascending, following the order of\;
  \Indp `bar' \(<\) `gridPos' \(<\) `pitch' \(=\) `rest' \(<\) `duration'

  \Indm\Indm \textit{tokens} \(\gets\) keep only the \textit{token} items for sorted \(T\).\;
  \KwRet{\textit{tokens}}

  \caption{\footnotesize {\sc Encode} encodes a melody to a token sequence.}
  \label{algo:encode}
\end{algorithm}

Notice that as soon as Positional Grid Tokens (e.g., \textsc{BAR}, \textsc{BEAT}, and \textsc{POS}) are encoded to the sequence, they introduce another time axis defined by themselves.
In the generation results of the current mainstream models, it is common to see inconsistency between the note-based accumulative time and the grid-indicated time,
which can be handled by different post-processing methods depending on the need of downstream tasks.
In this study, we only trust the note-based timing (from note/rest durations) and ignore positional grid tokens when decoding a generated token sequence, since the latter are only considered as auxiliary input helping a model to learn the temporal relationship.

\subsection{Model and Training Specifications}
Since the sequence length varies in encoding methods, the model should be good at handling long sequences.
Transformer-XL~\citep{daiTransformerXLAttentiveLanguage2019} is hence selected as it was the first model able to handle extra long sequences outperforming the LSTM network~\citep{hochreiterVanishingGradientProblem1998} and the vanilla transformer~\citep{vaswaniAttentionAllYou2017}.
Transformer-XL introduced the memory reuse mechanism and relative positional embedding to address the context fragmentation problem at the training stage, which is a influential improvement for music generation models.

However, we assume that the Transformer-XL of its original size (18 layers) for large text datasets is inappropriate for music generation task.
Since the vocabulary size is mostly from tens to hundreds, the model is very likely to be over-fitted according to our trials.
Hence, this study uses a 4-layer Transformer-XL, with 32 embedding dimensions and only 4 attention heads.
64 and 128 are used for the hidden dimensions and inner Feed-Forward layers, respectively.
As a result, this shrunk model only uses around 0.5M parameters, which is only 0.2\% of its original size.
It turns out that this tiny model could still be over-fitted for specific input representations, suggested by the continuously increasing test NLL loss soon after a few epochs.

An AdamW optimizer of learning rate 2e-4 is used as it contains regularization terms~\citep{loshchilovDecoupledWeightDecay2019}.
All the models are trained for around 25k steps (50 epochs) on the training set of max sequence length 1,024, and are tested on the same set of melodies.
During training, augmentation was performed on the training data for each epoch, with random transpositions within 6 semitones upward and downward.
For sampling, top-\(k\) sampling of \(k=5\) is used, starting with the token representing the beginning of a bar.
128 melodies of 512 tokens are sampled, \textsc{Pad} tokens removed in post-processing.

\subsection{Evaluation Metrics and Distribution Similarity}
Only objective metrics are used in this study, mainly because
this study focuses on how low-level feature encoding methods influence the model performance instead of improving the generation quality of the entire system.
Also, although a small network is used to test the model performance, the generation quality varies widely from resulted model.
Based on the listening experience of the authors, there exists obvious failure cases for some of the resulted systems.
Hence, we believe that the objective evaluation metrics are already enough to distinguish the different quality, so we did not conduct any subjective analysis.

The selected objective metrics are in two groups, pitch-based and rhythm-based, respectively.
\vspace{-1.5em}
\paragraph{Pitch}
\begin{itemize}
  \item \textbf{MAI} Mean Absolute Interval, measures the average steepness of the notes in a melody.
  \item \textbf{H(P)} Pitch Entropy, entropy of all the pitches (in MIDI numbers) in a melody.
  \item \textbf{H(PC)} Pitch Class Entropy, entropy of the 12 pitch class choices in a melody.
  \item \textbf{SC} Scale Consistency, defined as the largest pitch-in-scale rate over all possible major and minor scales.
  \item \textbf{MSD} Major-scale-rate vector standard deviation, the standard deviation of the 12 pitch-in-scale rates.
\end{itemize}
\vspace{-1.5em}
\paragraph{Rhythm}
\begin{itemize}
  \item \textbf{MD} Mean Duration, average note duration of a melody.
  \item \textbf{H(D)} Duration Entropy, entropy of all the note durations in a melody.
  \item \textbf{GC} Groove Consistency, the average similarity (rhythmic hamming distance) of every 2 consecutive bars.
  \item \textbf{EBR} Empty Beat Rate, the rate of beats where no note is being played or held.
\end{itemize}
From the metrics above, H(P), H(PC), SC, GC and EBR are implemented in the library muspy~\citep{dongMusPyToolkitSymbolic2020} and were first proposed in
the c-RNN-GAN~\citep{mogrenCRNNGANContinuousRecurrent2016}, MuseGAN~\citep{dongMuseGANMultitrackSequential2017}
and Jazz Transformer~\citep{wuJazzTransformerFront2020}.
We also introduce MAI, MSD, MD and H(D) in this study to better describe the distribution of the low-level pitch and durational features.

128 melodies are sampled from both the test set (truncated to training data length) and each resulted model.
The 7 metrics above are calculated for all the melodies.
For each distribution, the p.d.f is approximated by Gaussian kernel density estimation (KDE), whose bandwidth is chosen according to Scott's rule of thumb\citep{scottMultivariateDensityEstimation1992}.
To avoid over-smoothing of a large bandwidth and avoid the too strong assumption of normal distribution by Scott's rule of thumb, the bandwidth is further divided by 4.
Finally, the Overlapping Area (OA) is adopted to measure the similarity between all the model distributions and the true distributions.
Hence, a higher and closer-to-1 similarity indicates better approximation to the true distribution.

\section{Pitch Encodings}\label{sec:Pitch}
\subsection{MIDI Number and Class-octave}\label{subsec:number_co}
The popular MIDI number representation of pitches are integers from 0 to 127.
In the context of word embeddings, such embeddings provides no prior knowledge about the relationship among pitches to the model.
However, the pitch class pitch octave encoding breaks down the pitch feature into two tokens, as the name suggested.
Consider a pitch of MIDI number \(p\), the pitch class and pitch octave can be written as \(\left( p \mod 12, \left\lfloor \frac{p}{12} \right\rfloor \right) \).

Even though the two encodings methods can be always converted between each other, the class-octave encoding provides explicit information about pitch relationship, especially across different octaves.
For instance, a list of pitches (C4, E4, G4, C5, E5, G5) are encoded as (p60, p64, p67, p72, p76, p79) using number encoding,
but as (C, o4, E, o4, G, o4, C, o5, E, o5, G, o5) in class-octave manner.
In this example, the latter encoding clearly shows what pitch classes are being used, despite the change of octave.

The class-octave encoding also has more transpositional invariance compared to the MIDI number encoding,
simply because that transposing a pitch slightly would almost not change the octave,
transpositions in multiples of octaves would also change the pitch class.
In this sense, the similarity of pitch in music is better preserved in the class-octave encoding in an explicit way.
Hence, it is expected to result in better pitch and pitch class distribution for the generated melodies.

\subsection{Comparison Results and Discussions}
\begin{figure}[htbp]
  \centering
  \includegraphics[width=\linewidth]{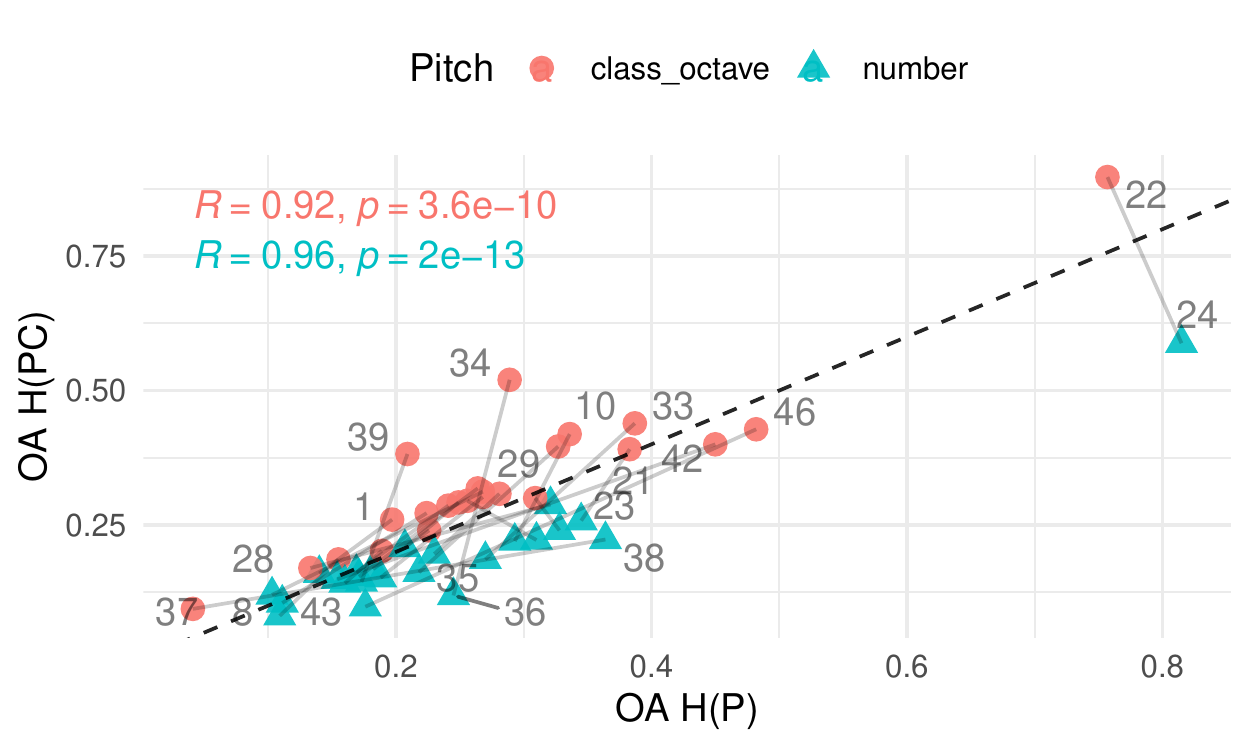}
  \caption{Resulted OA joint distribution of H(PC) and H(P). The dashed line represents \(y=x\), separating the two encoding methods. \textsc{Class-Octave} results are better in H(PC) while \textsc{Number} results are better at H(P), which is as expected.}
  \label{fig:hpc-hp}
\end{figure}

All the resulted 48 models of hyperparameter-grid can be grouped into 24 pairs that only differ in pitch options.
For the family of 7 selected metrics, a paired Wilcoxon test rejected 2 equal-mean null hypotheses, for MAI (\(p=\)6.57e-4) and H(PC) (\(p=\)4.22e-5), using the Holm–Bonferroni adjusted \(\alpha\) values controlling the family-wise error rate (FWER) \( \le 0.05\).
In terms of the gap, for MAI, the \textsc{Class-Octave} group sample distributions have around 0.158 higher OA compared to \textsc{Number};
for H(PC) it is 0.149 higher.
Another non-significant but worth-mentioning gap is 0.047 for H(P).
The overall distribution of H(P) OA and H(PC) OA are in Figure~\ref{fig:hpc-hp}.
The rest differences of OA are almost within the range of \(\pm\)0.02, which could be ignored.

\begin{figure*}[tphb]
  \centering
  \begin{subfigure}[b]{.32\textwidth}
    \centering
    \includegraphics[width=\linewidth]{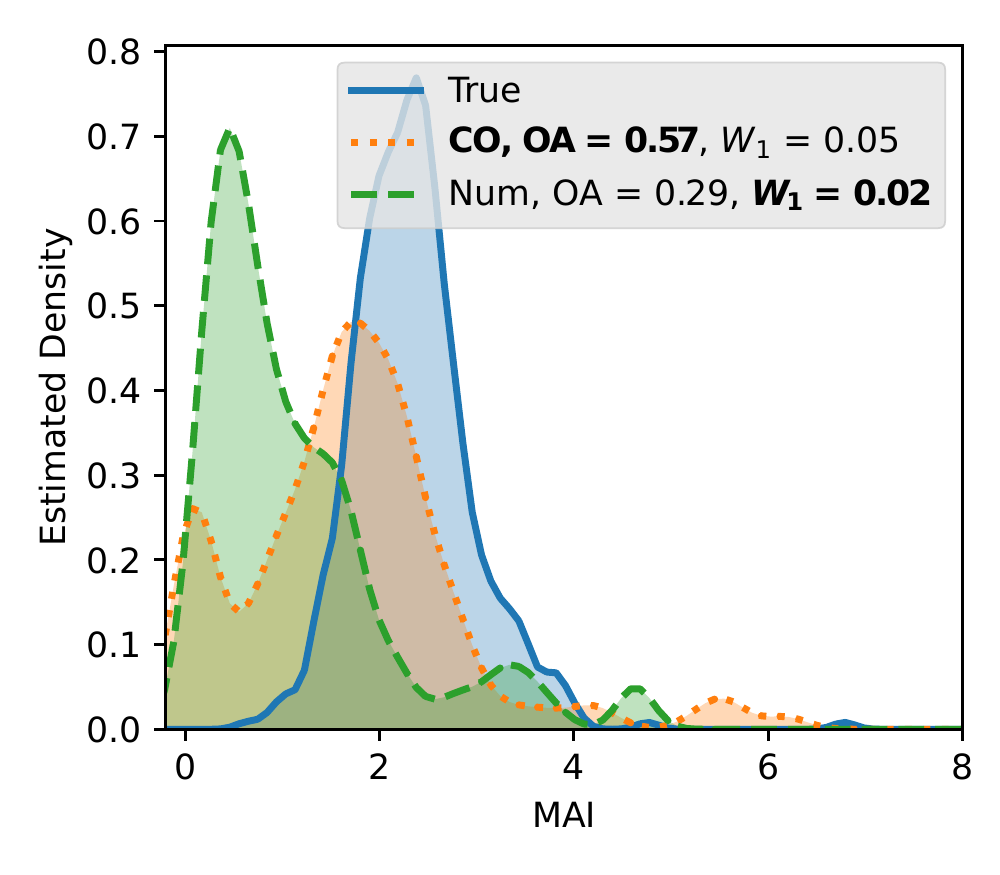}
    \vspace*{-2em}
    \caption{MAI distributions}
    \label{fig:pgd-MAI}
  \end{subfigure}
  \hfill
  \begin{subfigure}[b]{.32\textwidth}
    \centering
    \includegraphics[width=\linewidth]{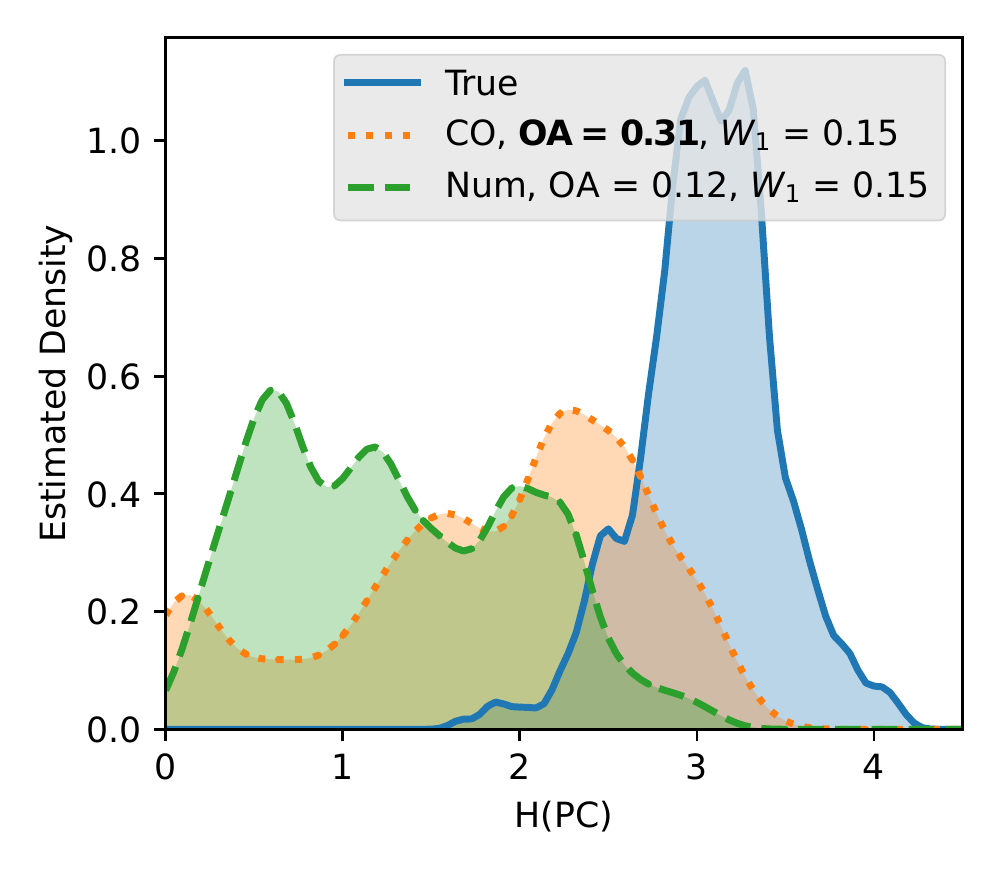}
    \vspace*{-2em}
    \caption{H(PC) distributions}
    \label{fig:pgd-hpc}
  \end{subfigure}
  \hfill
  \begin{subfigure}[b]{.32\textwidth}
    \centering
    \includegraphics[width=\linewidth]{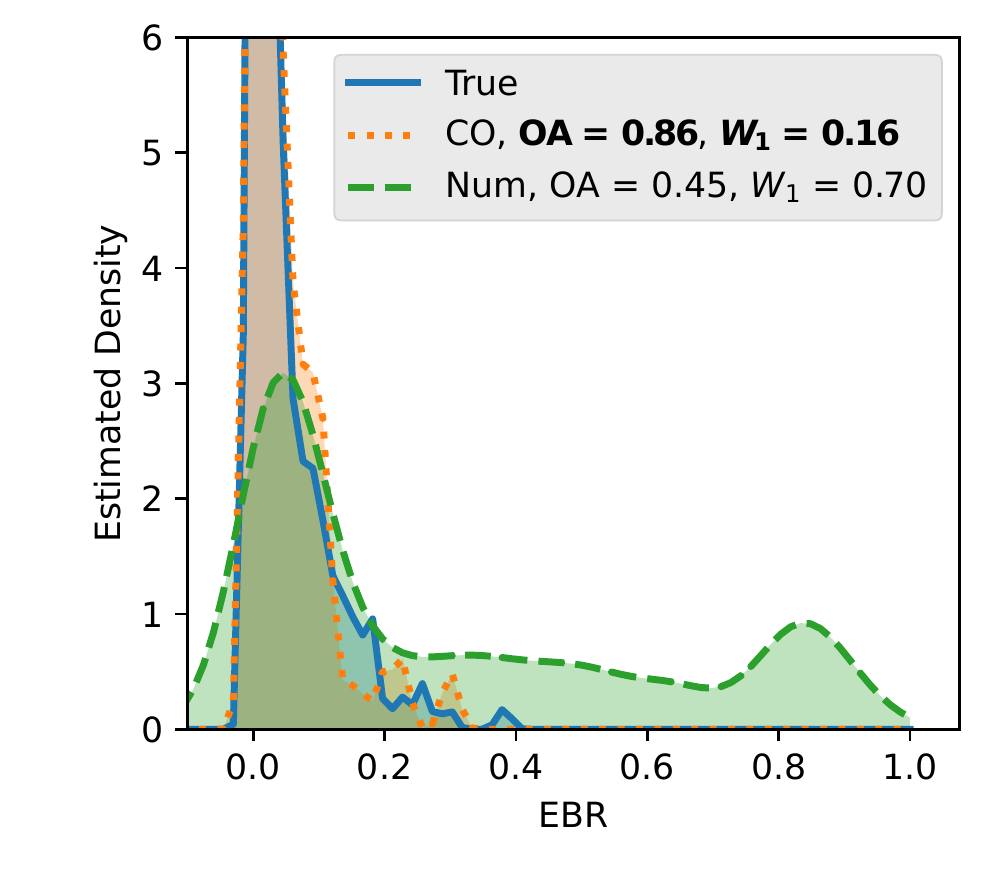}
    \vspace*{-2em}
    \caption{EBR distributions}
    \label{fig:pgd-ebr}
  \end{subfigure}
  \caption{Metric distributions for a representative pair of encoding (27, 28).
    \(OA\) and \(W_1\) denote the overlapping area and 1-Wasserstein distance, respectively, between the model sample distribution and the true distribution.
    \ref{fig:pgd-ebr} is a non-significant metric but also reveals salient differences of the two pitch encodings.}
  \label{fig:pitch_group_dists}
\end{figure*}

Around the mean area, we selected a representative pair of results (model 27 and 28) with relatively strong comparison, and plot the detailed distributions to observe their characteristics in Figure~\ref{fig:pitch_group_dists}.
Their corresponding encoding hyper-parameters are (PC = \textsc{Single}, PR=1, DR=16), with Pitch being \textsc{Number} and \textsc{Class-Octave}, respectively.

Figure~\ref{fig:pgd-MAI} shows that the \textsc{Class-Octave} model yielded a better MAI distribution which is less concentrated on 1 semitone and has higher density for the range of 1 to 4 semitones compared to the \textsc{Number} model.
In contrast, the \textsc{Number} is much more positively skewed, hinting that the generated melodies mostly progress in small intervals such as semitones and wholetones, which can be too conservative and non-exciting regarding the expected the listening experience.
The comparison result could be interpreted as the effectiveness of the separately encoded pitch classes and pitch octaves, despite its doubled length.
As mentioned before, the notes that are only a few semitones apart are most likely to share an octave token, or even two octave tokens differ by 1.
The similarity of pitches are better preserved and explicitly expressed on the token level.

It is also worth noticing that if the 1-Wasserstein distance metric is used instead to calculate the strict distances towards the true distribution, the \textsc{Number} is actually a closer Distribution of MAI.
However, here the OA focuses more on how much of the true distribution is being captured, so it is reasonable that a distribution of higher OA could actually be more distant.

The pitch class entropy H(PC) is another significantly better learnt distribution by the \textsc{Class-Octave} model. As in Figure~\ref{fig:pgd-hpc},
the mean entropy is around 2 to 3 bits, higher than the \textsc{Number} one.
An intuitive explanation for this could be that, in order to model the distribution of pitch classes, a \textsc{Number} model must
learn a meaningful pitch token space for all the 127 pitches, for instance, well clustered into 12 different pitch classes automatically,
or, follow a certain geometric pattern as constraints such that the model can rely on to generate correctly distributed pitches.
As a comparison, a \textsc{Class-Octave} model only has to model the relationship between the 12 given pitch class tokens that are known to be shared within the same octave.
In this case, the task is not from the scratch given the explicitly encoded constraints, thus less difficult.

From the perspective of data augmentation, random transpositions within a few semitones can be viewed as a kind of regularization, which permutes all the inter-pitch constraints in a cyclic group of 12 and resulted in only small changes of all the octave tokens.
For the \textsc{Number} token space, however, the changes do not happen on a cyclic group, which only explicitly benefits the neighbor pitches.
This tends to result in a smooth striped manifold of the pitch embedding space.
The analysis of the embedding space and the outliers would be in Section~\ref{sec:combined}.

Another interesting observation can be made on Figure~\ref{fig:pgd-ebr}, that the EBR distribution from the \textsc{Number} model has a heavy tail, which is not the case for the \textsc{Class-Octave} model nor the true distribution.

\section{Metrical Encodings}
\subsection{Position Complexity and Positional Resolution}
In recent studies such the REMI representation \citep{huangPopMusicTransformer2020},
the authors reported that the 16th note grid position produced the best results on the music generation task,
with several attempts of using other resolutions that resulted in worse performance.
To find out whether it is the sparsity or the absolute positions, or both that improved generation quality,
this study prefers a dense grid encoding, which encodes all the grid positions of a bar.
Importantly, all the encoded grid positions will be ignored during the decoding process,
since inconsistency handling is avoided, and we already have a special token \textsc{Rest} as a silent pitch token that fills the gaps between all the notes of a melody.

We define the hyper-parameter Position (bar) Resolution (PR) and Position Complexity (PC) of such encoding.
PR is a multiple of 4, indicating the number of even subdivisions of a bar of 4 beats.
PC has two options.
The option \textsc{Single} means the grid positions are all denoted by a single \textsc{Pos} token.
The bar line is provided by a \textsc{Bar} token before the first position.
Hence, by counting the occurrences after a \textsc{Bar} token it is able to calculate the offset to a bar.
The other option \textsc{Multiple} refers to separate absolute positions \textsc{Pos}\(_{1..PR}\) for a bar-level grid.
In this case, no separate \textsc{Bar} token is provided since the first absolute position necessarily means the beginning of a bar.

\begin{figure}[H]
  \begin{center}
    \includegraphics[width=.5\linewidth]{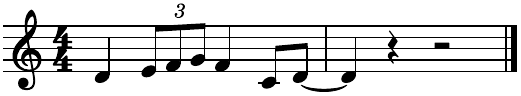}
  \end{center}
  \footnotesize
  \texttt{(Number, PC=\textbf{Single}, PR=4, DR=4) -> [\textbf{BAR} POS p62 d4 POS p64 d1 p65 d2 p67 d1 POS p65 d4 POS p60 d2 p62 d6 \textbf{BAR} POSx4]},
  length=24
  \vspace{.5em}

  \texttt{(Number, PC=\textbf{Multiple}, PR=4, DR=4) -> [POS0 p62 d4 POS1 p64 d1 p65 d2 p67 d1 POS2 p65 d4 POS3 p60 d2 p62 d6 POS0..3]},
  length=22
  \vspace{.5em}

  \texttt{(Number, PC=\textbf{Single}, PR=16, DR=4) -> [\textbf{BAR} POS p62 d4 POSx4 p64 d1 POS p65 d2 POSx2 p67 d1 POS p65 d4 POSx4 p60 d2 POSx2 p62 d6 POS \textbf{BAR} POSx16]}, length=48
  \vspace{.5em}

  \texttt{(Number, PC=\textbf{Multiple}, PR=16, DR=4) -> [POS0 p62 d4 POS1..4 p64 d1 POS5 p65 d2 POS6..7 p67 d1 POS8 p65 d4 POS9..12 p60 d2 POS13..14 p62 d6 POS15 POS0..15]}, length=46
  \caption{An example of the same melody encoded with different PC and PR settings}
  \label{fig:licc}
\end{figure}

As an example, for the melody in Figure \ref{fig:licc}, different PC and PR settings can yield the following encoded sequences.
Pitch=\textsc{Number}, PR=4 (4 downbeats) and PR=16 (all the 16th notes) are for elaboration.
We use a conciser notations to shorten the sequence length on the paper: \texttt{<TOKEN>x}\(n\) denote a token repeated by \(n\)-times and \texttt{<TOKEN>}\(a..b\) represents a range of tokens.

The first benefit of this design is the similarity between \textsc{Single} and the \textsc{Hold} token used by DeepBach~\citep{hadjeresDeepBachSteerableModel2017}, but also different from DeepBach that this repetitive dense grid provides explicit bar lines and it does not determine any note duration at all,
so that we can investigate whether such repetitive grid helps modeling metrical features.
Also, the dense setting allows the comparison between \textsl{Single} and \textsc{Multiple} to be conducted with almost equally long encoded sequences.

It is important to ensure similar sequence lengths when comparing metrical encodings, since the transformer
model used both in the REMI work and ours are trained with teacher-forcing and non-weighted NLL loss.
This means that for every batch gradient update, the token-wise average loss is weighted according to the frequencies of different token types in the batch, thus determining the learning priorities.
For example, provided that a melody is encoded into a longer sequence \(A\) with a large amount of grid tokens, and a shorter sequence \(B\) with sparsely encoded absolute grid positions.
When the loss is averaged along steps, the losses for grid token steps are more weighted in \(A\) than in \(B\), so the optimization direction will lean more towards the positional tokens because of \(A\)'s encoding.

In the experiment settings, both PC options are considered.
PR = (0, 1, 4, finest) are compared, where
PR = 0 denotes the ablated group that does not use the bar-level position grid feature at all (the PC being undefined of course), and
the \texttt{finest} is calculated by DR \(\times\) 4, covering 16, 32, 48 and 64.
In the ablated group there are 8 models and the control group there are 40 models.

\subsection{Results and Discussion}
In this subsection, the results are compared in three ways: ablation study, PC and PR.

\subsubsection{Ablation Study}
Among the family of 9 metrics, Wilcoxon tests resulted in two relatively significant differences of metric distribution OA for the ablated group and the control group.
The null hypothesis is that the two groups share the same OA in for all the 9 metrics and is tested according to the Holm–Bonferroni method.
At a FWER no greater than 5\%, we failed to reject the null hypothesis.
However, there are two OA difference that are with small \(p\)-values which deserves discussion:
the control group has higher average OA for Mean Duration (MD) at \(p_1 = 0.0059\), slightly greater than \(\alpha_1 = 0.0055\),
and higher average OA for Duration Entropy (H(D)) at \(p_2 = 0.0089\), slightly greater than \(\alpha_2 = 0.0063\).
The box plots are in Figure ~\ref{fig:pos_comp-md,hd}.

\begin{figure}[phtb]
  \centering
  \includegraphics[width=.7\linewidth]{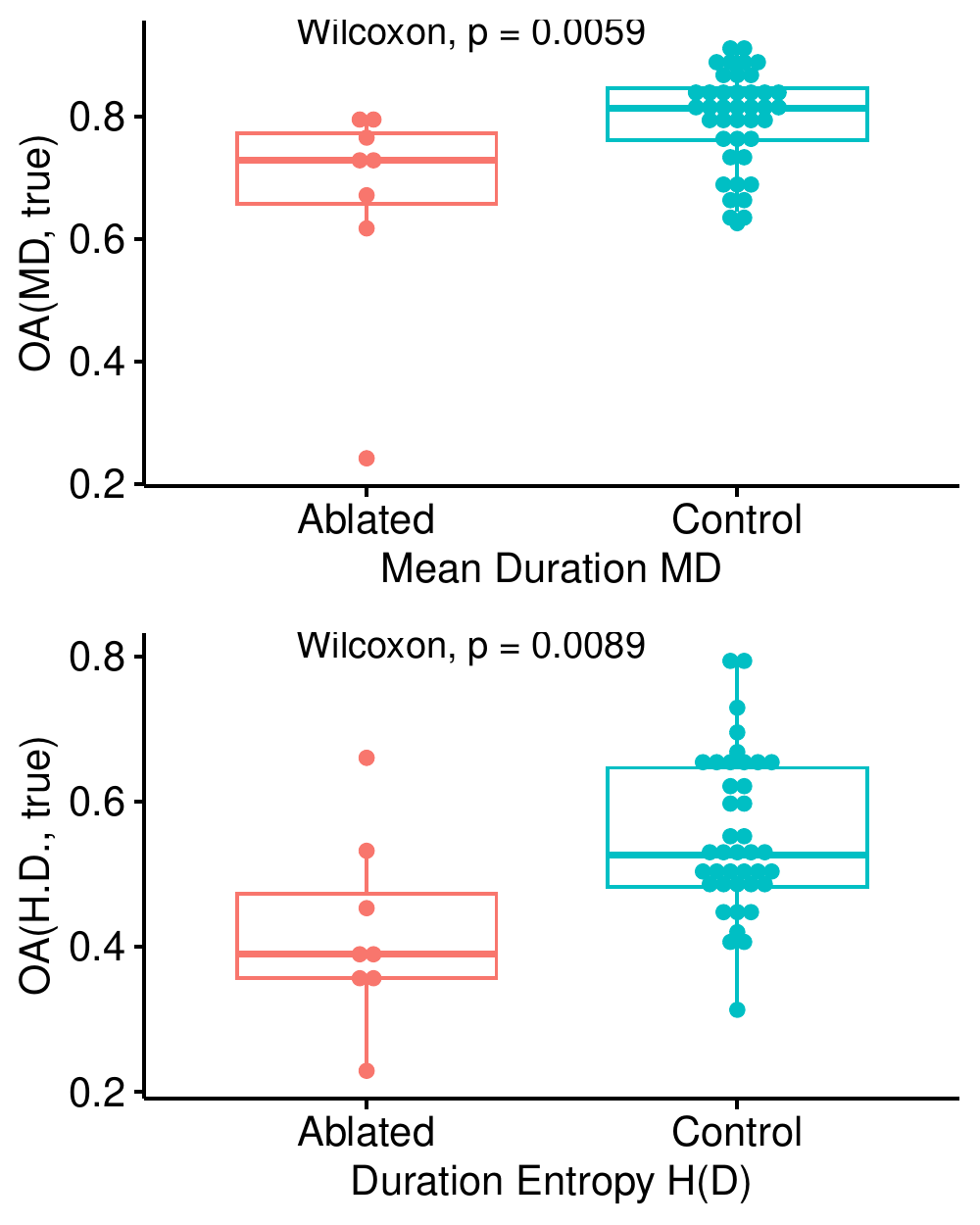}
  \caption{Metric distribution OA of Ablated group and the control group}
  \label{fig:pos_comp-md,hd}
\end{figure}

Among all the metrics, the two noticeably improved metrics are both about the distribution of note duration,
even if the position grid does not determine the note durations.
This possibly suggests that the grid features being helpful during the learning of duration features.
Without the grid, the only approach to describing the note onsets and offsets are by accumulating the duration tokens (from either pitches or \textsl{REST}).
When the grid is provided, the relative position from the bar line can be an additional information source to modelling the note durations.
This result also matches with the feasibility of REMI's sparse encoding of tokens.

\begin{figure}[phtb]
  \centering
  \includegraphics[width=.7\linewidth]{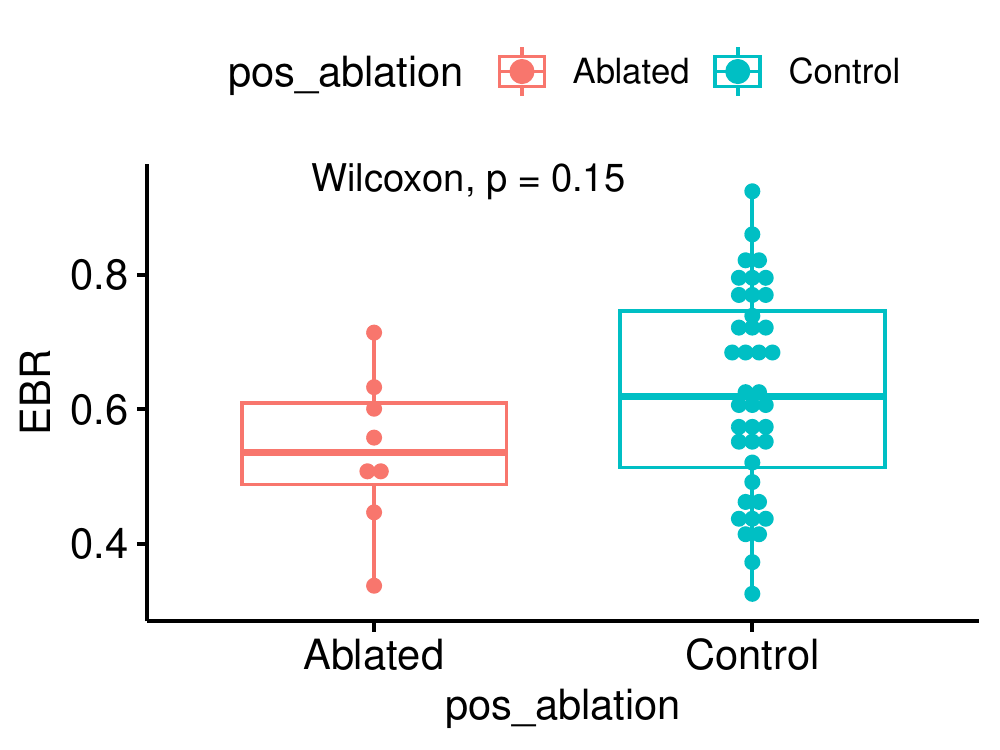}
  \caption{When the positional grid feature is encoded, higher OA of EBR distribution is achieved.}
  \label{fig:pos_comp-ebr}
\end{figure}

Another non-significant metric, Empty Beat Rate (EBR) has unadjusted the \(p\)-value of 0.15, but the ranges of the distributions are worth a plot,
see Figure~\ref{fig:pos_comp-ebr}

The rest metric OAs are either slightly increased for the control group or similar in distribution, which will be skipped.

\subsubsection{Interaction of Position Complexity and Position Resolution}
If only grouped by PC, 32 out of the 40 models with PR \(>\) 1 can be grouped to 16 pairs only different in PC.
A paired Wilcoxon test at FWER no greater than 0.05 failed to reject the null hypothesis, meaning there is no significant difference on the group mean.
The grouped box plots showed that the influence of PC varies with PR and Pitch encoding, which would be analyzed in Section~\ref{sec:combined}

\begin{figure*}[hbtp]
  \centering
  \begin{subfigure}[t]{\linewidth}
    \centering
    \includegraphics[width=\linewidth]{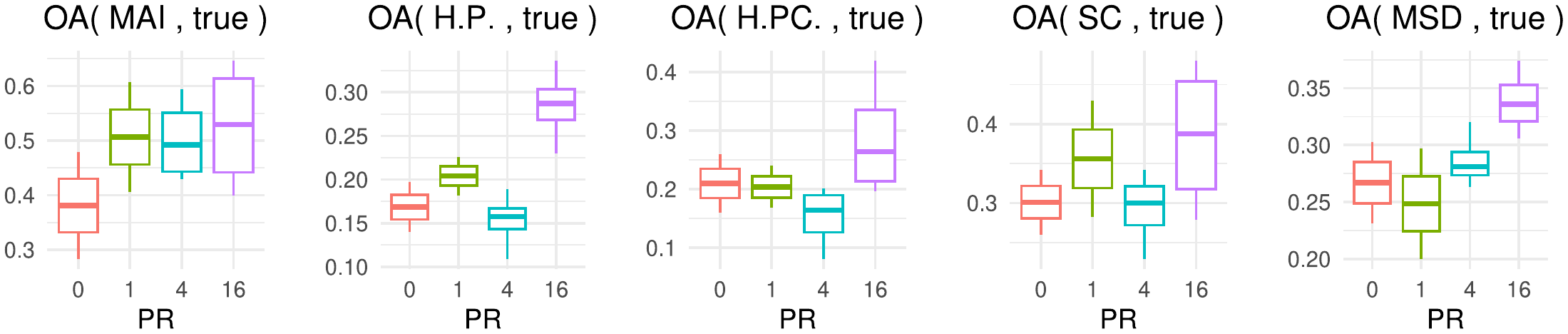}
    \caption{Pitch-related features}
    \label{subfig:pr-5}
  \end{subfigure}

  \vspace{1em}

  \begin{subfigure}[b]{\linewidth}
    \centering
    \includegraphics[width=.8\linewidth]{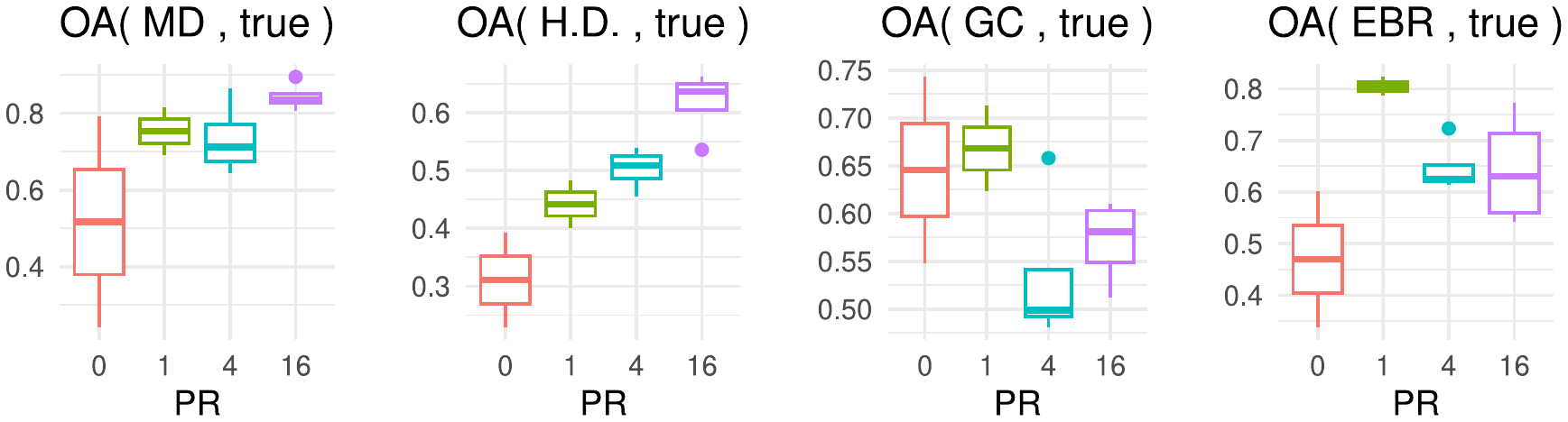}
    \caption{Rhythmic features}
    \label{subfig:pr-4}
  \end{subfigure}
  \caption{OA distributions as PR increased. Higher values are better. O for the ablated group, 1 for bar, 4 four downbeats, 16 for the finest grid at DR = 4.}
  \label{fig:PRs}
\end{figure*}

Given DR = 4, we gathered 12 models that could differ in other settings, with PR options of: O, ablated; 1, only \textsc{BAR}; 4, only downbeats; 16, the finest grid under the DR = 4.
Figure~\ref{fig:PRs} plots the OA of different PR, with \ref{subfig:pr-5} about the 5 pitch-related metrics and \ref{subfig:pr-4} about the 4 rhythmic metrics.

Although it is designed as a smooth transition from PR = 0 to PR = 16, the results are not necessarily smoothly interpolated as expected.
Three observations on the trend of OA against PR are made on the results.

\paragraph{Observation 1} Among all the metrics, the two most benefited metrics are MD and H(D), the two regarding note durations, since they both increased from a poor value to more than 0.7, which indicates a relatively good approximation.
These two metrics also display a stabler increase with smaller variances compared to other metrics.

\paragraph{Observation 2} Among the 5 pitch-related metrics, most fluctuate when PR = 1 and 4.
The only prominent improvements happen at PR = 16 but are also not much.
Given the relatively small sample sizes and small ranges of OA, the fluctuations among these features can be ignored.

\paragraph{Observation 3}
It can be noticed that the EBR reaches the best up to 0.8 when PR = 1, i.e., only additional \textsc{Bar} tokens are add,
and deteriorates as PR increases, with the smallest step shorter than a beat.
GC, on the other hand, show a decreasing trend.

Regarding observation 1, the improving note duration distribution as PR increases seems to indicate the engagement of such grid as an alternative way to help estimate a reasonable note duration in a melody, which also to some extents implies that the model is relying less on accumulating previous duration tokens for an absolute time.

If this conjecture holds, observation 3 can also be explained in a similar way.
Since the model relies more on the grid position to estimate the durations,
it could be less attentive to the durations of the previous notes.
Also, in the decoding procedure, the grid is not used to correct the durational inaccuracies,
so they are accumulated and amplified,
resulting in a worser beat-wise EBR as in the plot, let alone the bar-wise GC which is even worser than the ablated group.

To summarize, the results reveal a compromise between using the additional metric grid or the accumulating duration.
When the imposed grid is more focused (with higher PR, increasing the proportion of grid tokens), the distributions of durations are better modeled.
However, the quality of beat-level and bar-level groove seems correspondingly decreased.
The appropriate PR to reach the subtle balance seems to vary in specific metrics.

\section{Durational Resolution}

Durational Resolutions (DR) is an alias for the terminology Ticks-Per-Quarter-Note (TPQN), as used in the MIDI specifications,
refers to the number of subdivisions a quarter note, using one as the unit which all the time spans are a multiple of.
Here, DR is dedicated to note durations so that it is independent from PR.

The process of discretizing all the note durations into multiples of DR usually causes information loss such as tuplets.
One way, as used in this study, is to first round the onsets and offsets of all the notes and then calculate the note durations.
For example, suppose DR is 4, 8th note triplet can have a duration of 1 or 2 depending on the onset.
The example can also be seen in Figure~\ref{fig:licc}, where the second to the fourth notes turn out to have durations 1, 2 and 1, respectively.

The problem of an unreasonably low DR is obvious because too much information is lost for reconstruction.
A large DR, on the other hand, tends to increase the model's learning difficulty because the tiny subtle numeric differences must be
learned to create reasonable combinations that add up in time.

In this section, the 12 models with PR = \texttt{finest} are compared on different DR settings.
That is, the DRs are set to (4, 8, 12, 16) and the corresponded PRs are (16, 32, 48, 64).
They vary in two Pitch options and two PC options.
The PR = \texttt{finest} is chosen since the previous experiments before have shown that most metrics are improved at the finest PR.

\subsection{Representative results}\label{subsec:DR-results}
\begin{figure*}[hbtp]
  \centering
  \begin{subfigure}[t]{\linewidth}
    \centering
    \includegraphics[width=\linewidth]{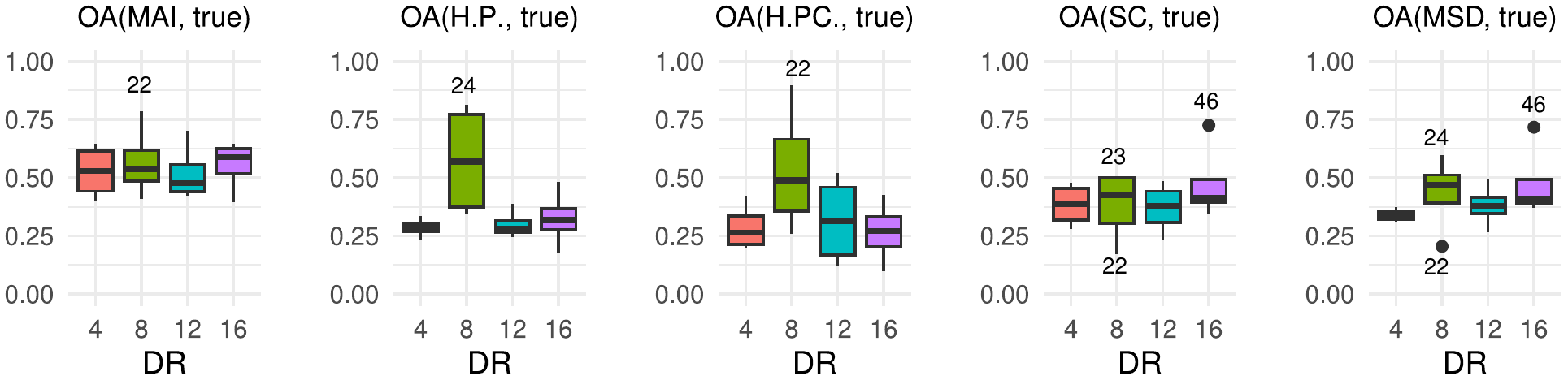}
    \caption{Pitch-related features}
    \label{subfig:dr-5}
  \end{subfigure}

  \vspace{1em}

  \begin{subfigure}[b]{\linewidth}
    \centering
    \includegraphics[width=.8\linewidth]{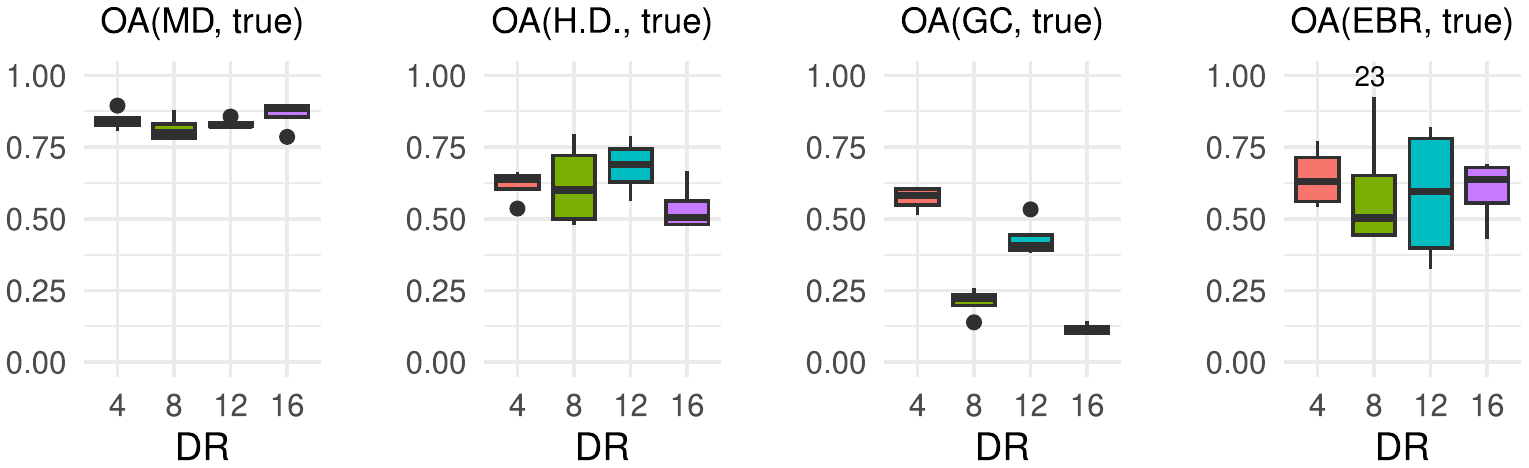}
    \caption{Rhythmic features}
    \label{subfig:dr-4}
  \end{subfigure}
  \caption{OA distributions as PR increased. Higher values are better. O for the ablated group, 1 for bar, 4 four downbeats, 16 for the finest grid at DR = 4.}
  \label{fig:DRs}
\end{figure*}

Results do not show a simple linear relationship and are not well fitted using multivariate multiple linear regression,
hence they are plotted in Figure~\ref{fig:DRs} and discussed in groups.

The general observed trend for the 5 pitch-related metric OAs is that they are slightly improved as the DR increases; but the GC OA quickly dropped.
Extreme cases are noticed at DR = 8 where PR = 32, for the 4 models (id = 21 to 24), with quite a few metrics noticeably high.
The corresponding obvious outliers are annotated as model IDs in the plots.
If the neighbor configurations such as DR = 4 and DR = 12 are also taken into consideration, the DR = 8 group seems to pulling the the neighbors' performance towards it, probably indicating an non-monotonic influence of PR with a peak at DR = 8.
After checking, the two outlier models (22 and 24) also contributed to the maxima of MD, H(D) and the minima for the plots regarding GC and EBR in Figure~\ref{subfig:pr-4}.
Since the overall trend of MD and H(D) is subtle, we will ignore this two items for this special case.
However, in contrast to the best performance of approximating pitch-related metrics, model 22 and 24 on the other hand learned very poorly about grooves and beats.
The large variances in the DR = 8 group is mainly from different Pitch and PC options, whose interactions with DR would be analyzed with more details in Section~\ref{sec:combined}.

The results above suggest that DR, similar to PR, also has a non-monotonic influence on the metrics.
Neither too low or too high DR results in optimal approximation to the true metric distributions.
The opposite conditions for the extreme cases also indicate that the optimal PR can differ in metrics.
In our case, the optimal DR seems to be 8 or 12, which is consistent with the optimal PR 16, as reported in the REMI's original work~\citep{huangPopMusicTransformer2020} and 24,
as reported in the piano-roll MuseGAN \citep{dongMuseGANMultitrackSequential2017}.
This could also be due to a high DR causing the model being over-fitted to the training dataset, which would be discussed in Section~\ref{sec:combined}.

\section{Combination Analysis}\label{sec:combined}
From the previous experiments we have observed two phenomena.
First, the impact of two resolutions PR and DR are both non-monotonic, and the optimal DR and PR even vary in metrics,
e.g., best approximated MD and H(D) are reached at a high PR, while a better approximated GC have low PR.
This suggests a kind of trade-off of the model performance on different metrics.
Second, the outliers in a group are sometimes caused by both 2 options of the \textbf{other} hyper-parameters,
which further hints the interaction of the hyper-parameters.

In this section, we will discuss two stages of the music generation task where additional factors apart from the encoding hyper-parameters must be taken into consideration to explain the trade-off:
the task goal and the learning process---the stage after encoding, and the data quality---before the encoding.

\subsection{Position Complexity and the Exposure Bias}
\begin{figure}[htbp]
  \centering
  \includegraphics[width=\linewidth]{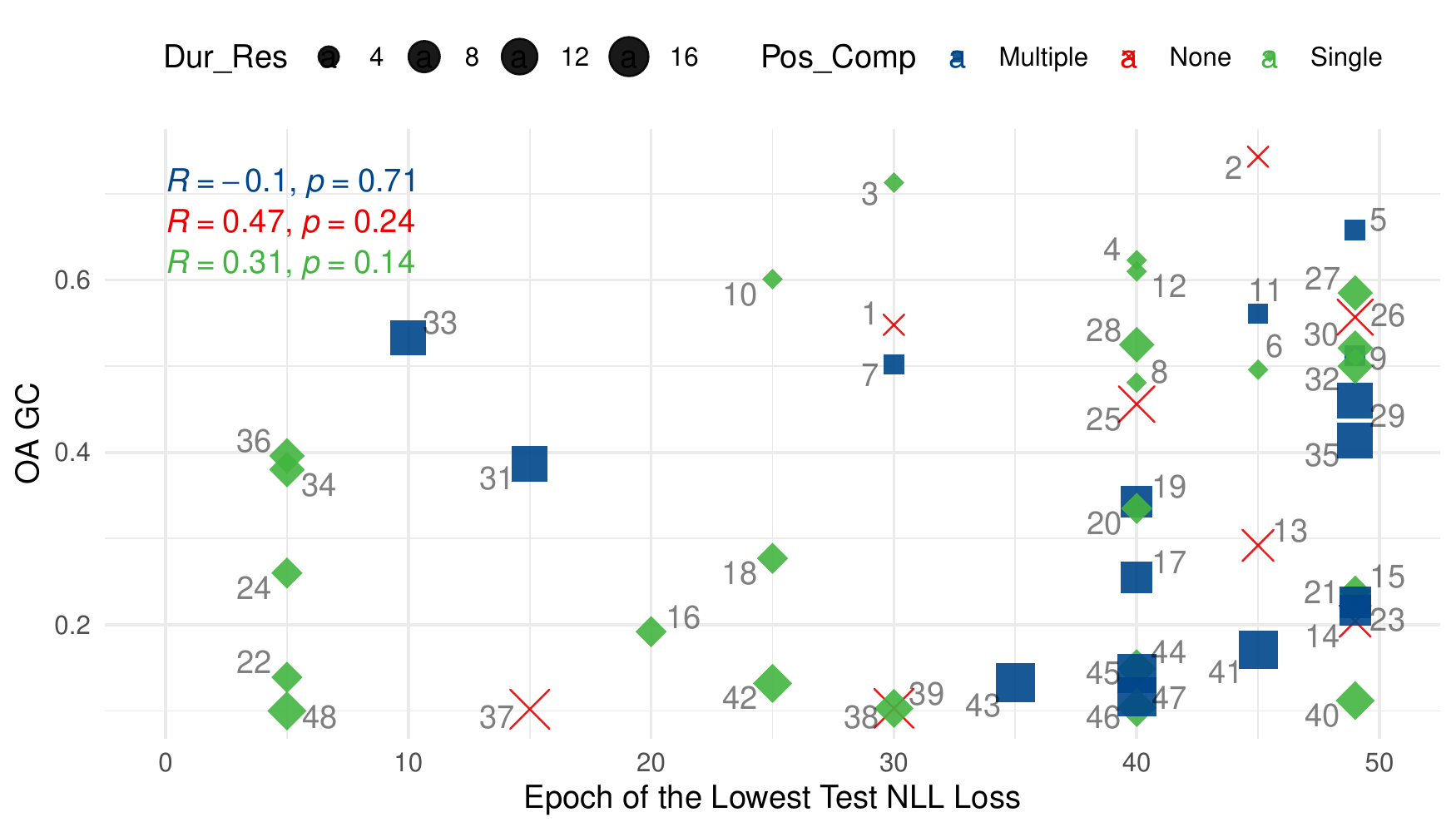}
  \caption{OA of GC against the best epoch, colored by Position Complexity.}
  \label{fig:best_epoch-GC}
\end{figure}

During the training process, we have noticed that, the test loss of some models started to keep increasing till the end of all the 50 epochs.
All the experimented models are evaluated at the epoch with the lowest test loss (the closest checkpoint is used).
The groove-related GC metric is used to reveal the relationship between metric approximation performance and the best epoch, plotted in Figure~\ref{fig:best_epoch-GC}.

Figure \ref{fig:best_epoch-GC} shows a trend of better GC approximation after trained longer to reach the lowest loss.
Especially for \textsc{Single} group, the OAs at the initial epochs are poor.
As the number of training steps increases, the \textsc{Single} group starts to have performance gain while the \textsl{Multiple} group becomes worse and stays at a low level around 0.2.

Another interpretation on this plot is that as PR and DR increases, (plotted with increasingly larger marker sizes), the models shifted from the upper right (slower training, higher performance) to lower (for \textsc{Multiple}, longer training but worse) and lefter (for \textsc{Single}, early convergence with test loss never becomes lower, models collapsed)

To summarize Figure~\ref{fig:best_epoch-GC}, smaller PR and smaller DR resulted in more consistent grooves.
\textsc{Single} tokens should not be learnt too fast that cause an over-confident model failure.

We believe that this is caused by the auto-regressive nature of the task, with teacher forcing used to speed up the convergence and correct errors in the early stage.
When discussing the learning process of a Transformer-based music generation model,
It is sometimes, not frequently, mentioned that when teacher forcing is applied, the averaged cross-entropy loss is equivalent to maximizing the log-likelihood of the input sequence~\citep{schmidtGeneralizationGenerationCloser2019}.
For the sequence, by keep applying the chain rule to the conditional probability, the model likelihood \(p(x_{1:n})\) can be expanded into products of step predictions \(p(x_n|x_{1:n-1})\), namely,
\[
  \begin{split}
    p(x_{1:n}) &= \prod_{k} p(x_k|x_{1:k-1})\\
    \log p(x_{1:n}) &= \sum_{k} \log  p(x_k|x_{1:k-1})
  \end{split}
\]
The mean negative log terms are the cross entropy loss between the predicted step logits over the vocabulary distribution and the one-hot labels.
A common problem of teacher forcing is the exposure bias, i.e., the discrepancy between the high likelihood of training samples and worse generated qualities or model over-fitting, which is observed in the experiments.

The Maximum-likelihood nature also makes the loss sensitive to the true distribution.
In our case, as the PR increases, \textsc{Single} encoding of metric grid results in highly repetitive tokens in the training sequence, which accounts for a large proportion of the step-wise averaged loss.
The problem can be addressed by scheduled sampling~\citep{mihaylovaScheduledSamplingTransformers2019}, or using weights to balance the token in the vocabulary, with the help of domain knowledge~\citep{schmidtGeneralizationGenerationCloser2019}.
However, as the author stated, this approach is usually not computationally efficient and in our case it can also require tedious turning process of the weights.

\begin{figure}[hpbt]
  \centering
  \includegraphics[width=\linewidth]{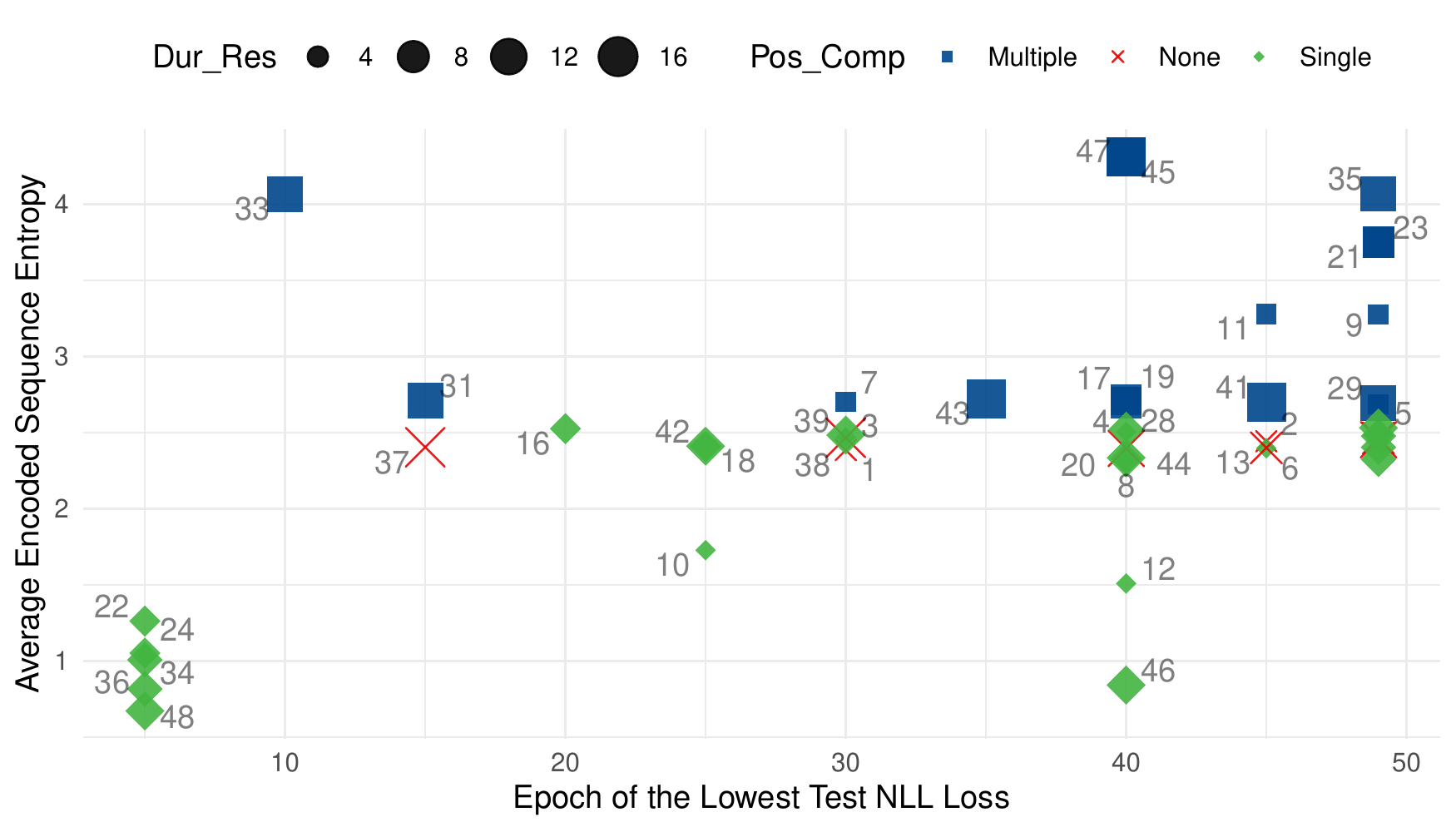}
  \caption{Diverged performance of PR under with different encoding entropy}
  \label{fig:label_entropy}
\end{figure}

The differences of \textsc{Single} and \textsc{Multiple} could also be interpreted from the angle of the entropy of encoded sequences.
For a higher PR, the repetitive single \textsc{Pos} tokens decreases the entropy of the true sequences while the multiple absolute
grid tokens, appearing with equal frequencies, increases the entropy,
which resulted in diverged task difficulty (of minimizing the loss), see Figure~\ref{fig:label_entropy}.

\subsection{Pitch Embedding Space Over-fitting and Data Quality}

\begin{figure}[hpbt]
  \centering
  \includegraphics[width=\linewidth]{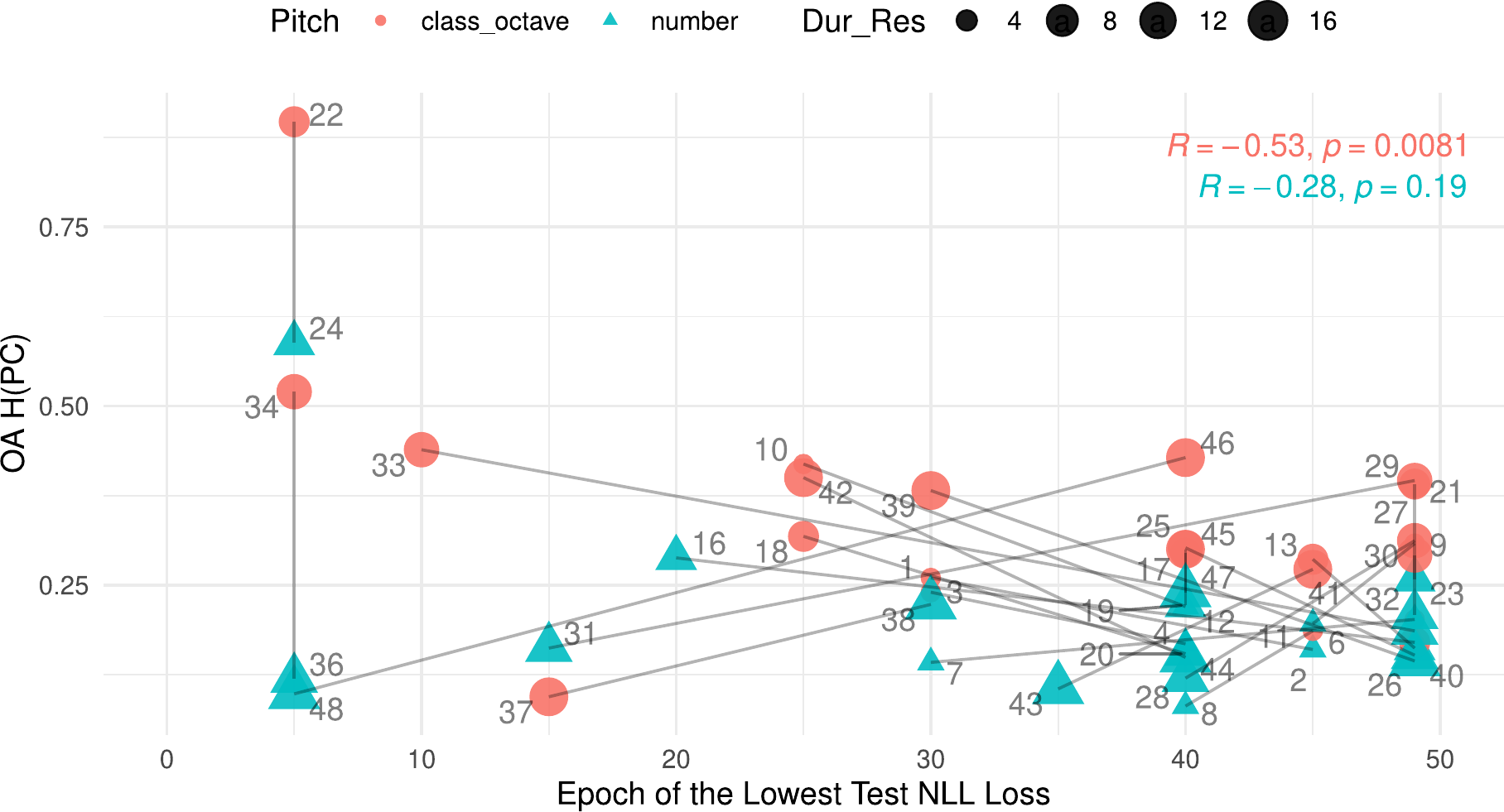}
  \caption{OA of H(PC) against the best epoch, grouped by Pitch option.}
  \label{fig:best_epoch}
\end{figure}

From Figure~\ref{fig:best_epoch} two the \textsc{Class-Octave} group is prominently better than the \textsc{Number} group.
However, the lower plot shows that the models have much worse approximations of the pitch classes if trained longer. The best models (model 22, 24 and 34) are all from the epoch 5, which mostly because of the PC = \textsc{Single} is used with a large PR.

The early stopping caused by other hyper-parameters brings out the decent pitch-related modeling.
Fortunately, the pitch embedding space can be checked and compared through dimension reduction and be visualized.
Hence, we choose model 22, 24, 37 and 8, the four extreme cases to compare the differences.

\begin{figure}[hpbt]
  \centering
  \begin{subfigure}[t]{.45\linewidth}
    \centering
    \includegraphics[width=\linewidth]{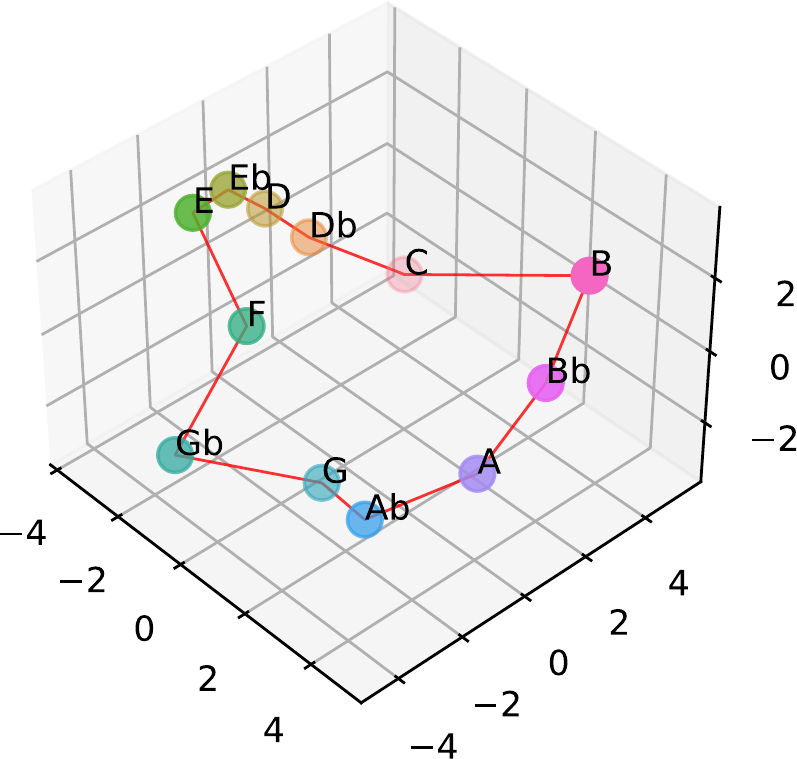}
    \caption{22-\textsc{Class-Octave}}
    \label{fig:emb-22}
  \end{subfigure}
  \hfil
  \begin{subfigure}[t]{.45\linewidth}
    \centering
    \includegraphics[width=\linewidth]{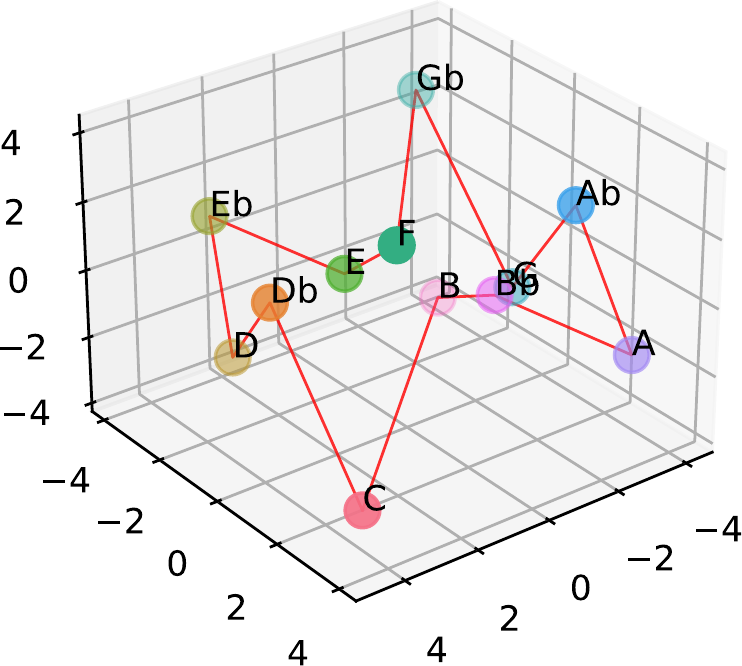}
    \caption{37-\textsc{Class-Octave}}
    \label{fig:emb-37}
  \end{subfigure}
  \vfill
  \begin{subfigure}[t]{.45\linewidth}
    \centering
    \includegraphics[width=\linewidth]{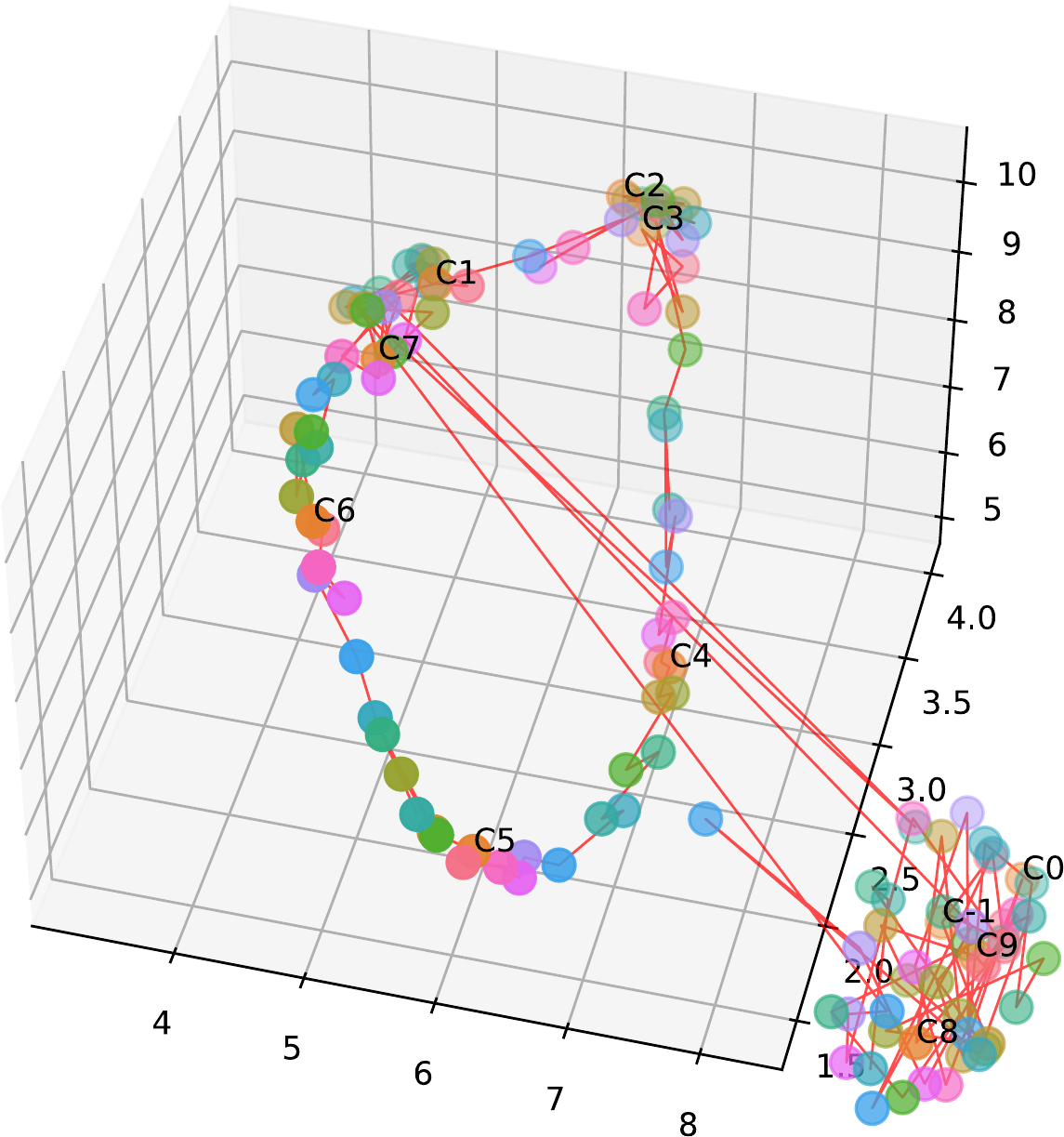}
    \caption{24-\textsc{Number}}
    \label{fig:emb-24}
  \end{subfigure}
  \hfil
  \begin{subfigure}[t]{.5\linewidth}
    \centering
    \includegraphics[width=\linewidth]{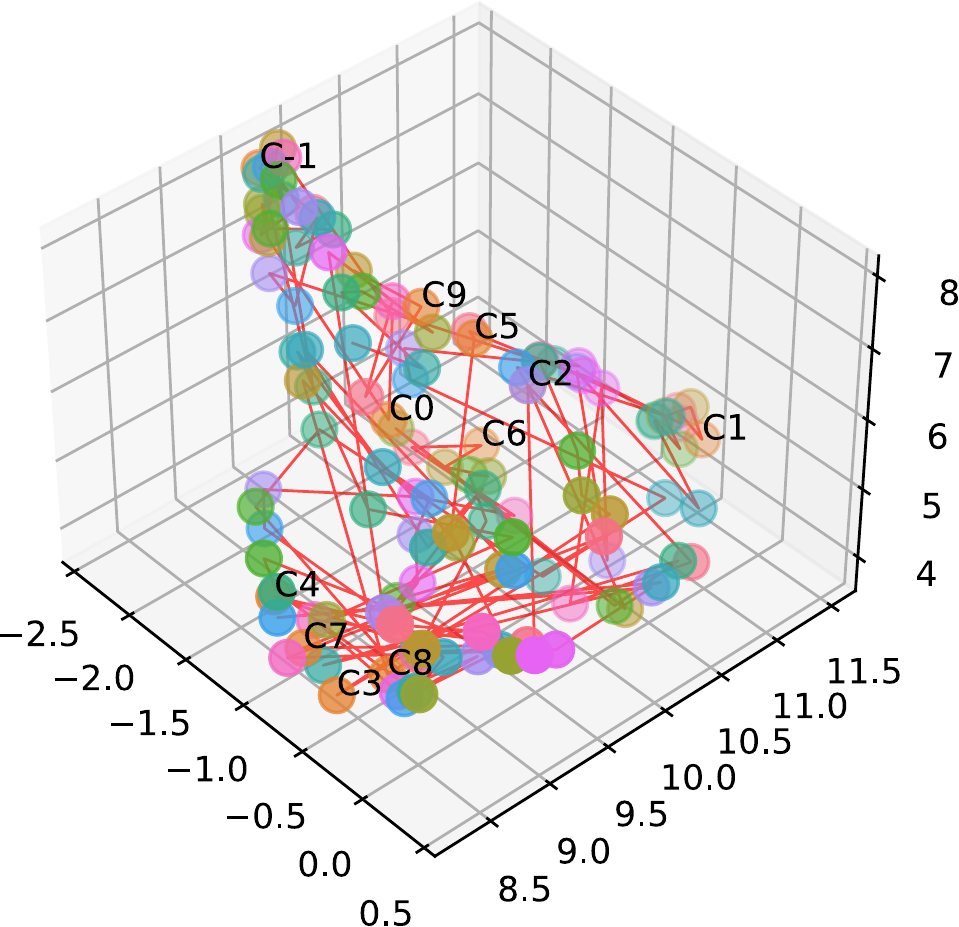}
    \caption{8-\textsc{Number}}
    \label{fig:emb-8}
  \end{subfigure}
  \caption{Extreme cases of the embedding space. The two best cases are on the left, and they both come from an early stopped model.
    The upper two are reduced from 32-dimensions using principle component analysis (PCA) while the lower two are obtained by uniform manifold approximation and projection (UMAP) to avoid crowdedness.
  }
  \label{fig:emb}
\end{figure}

From Figure~\ref{fig:emb}, the early-stopped models, also the two models of closest H(PC) to the test set, have much smoother pitch embedding spaces than those being trained for a many epochs.
Also, the clear relationship of the pitch classes \ref{fig:emb-22} and that of pitches \ref{fig:emb-24} matches with the expected striped manifold\footnote{
  such manifold is also visualized in the literature such as the PianoTree VAE \citep{wangPIANOTREEVAEStructured2020}
},
which means the embedding spaces have already modeled the proximity of adjacent pitches, especially benefited by random transposition of melodies at the training stage.
The problem displayed by the worst two, suggest that the further training breaks such relationship because modelling the noises in the dataset becomes more important in minimizing the NLL.
Without such smooth pitch relationship, the generated sequences hence do not show closer distribution by any means.

\begin{figure}[htbp]
  \centering
  \includegraphics[width=.6\linewidth]{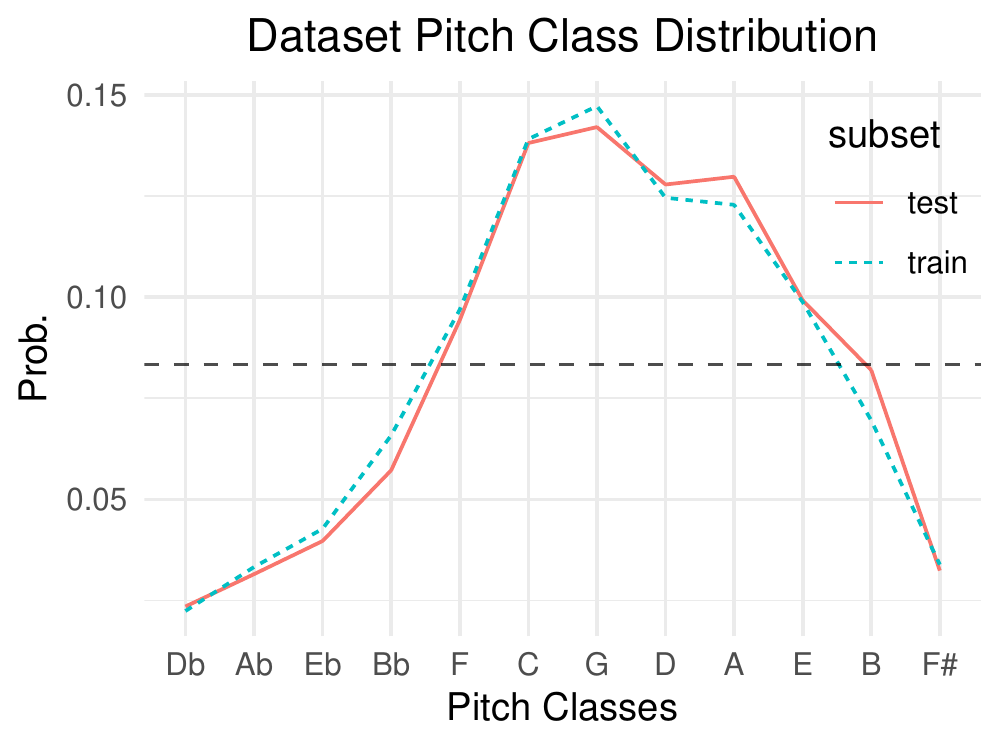}
  \caption{The pitch class distributions of all the raw samples in both the training set and the test set. Such distribution is formed because most samples in the dataset are in the C major key or a minor key.}
  \label{fig:pc-dist}
\end{figure}

Another indicator of the pitch embedding space being well-fit in the early stage is that,
the (visualized) embedding spaces (\ref{fig:emb-22}) already hints a pitch class distribution that is biased to that of the training dataset,
which is plotted in Figure~\ref{fig:pc-dist}.
Since the true distribution mostly features the notes in the C major scale, the embedding space also shows some irregularity and is twisted to fit the true distribution.
In comparison, the pitch classes in \ref{fig:emb-37} shows a much worse over-fitted situation---the ``black keys'' (D\(\flat\), E\(\flat\), G\(\flat\), A\(\flat\), and B\(\flat\)) are noticeably extruding out,
with the remaining (C, D, E, F, G, A, B) lying on a lower surface, which is exactly the distribution of notes of the dataset,
so the diversity of the generation system is affected.

The \textsc{Number} embedding space \ref{fig:emb-24} suggests that, even for the top-performing model, in the corner there are a cluster of rare pitches.
The similar situation for the \textsc{Class-octave} space is that the outliers are octave tokens such as \texttt{'o0'}, \texttt{'o1'} and \texttt{'o8'}, \texttt{'o9'}, but it is not worth a plot.
In the more over-trained \ref{fig:emb-8} adjacent pitches are almost in distinguishable by their locations, which are not advanced patterns but the noise.
The over-fitting problem seems to be worse for \textsc{Number} pitch encoding since they use more vectors, i.e., more parameters to model the pitch relationship.

To summarize, even in the low dimension of 32, the pitch embeddings can show satisfying approximation of the true distribution,
suggesting the unnecessariness of an unreasonably high dimensions such as 512.
Also, such low dimensional embedding still suffers from the problem of easy over-fitting,
which leads to the problem of rethinking about the effectiveness of early works, where static vector representations of pitches were used in rule-based systems with both satisfying results (in terms of pitch and pitch class),
and even stronger interpretability.

Quite different from a natural language with a large vocabulary,
the pitch relationship is based on a much smaller set of units, e.g., only 12 pitch classes,
and can be shared across different cultures as universally recognized.
Therefore, an explainable and semantic representation should preferred.

For the practical recommendations of symbolic music tasks, such as generation,
we argue that the input pitch representation should be better designed as a pre-determined, domain-knowledge-based, algorithmically-extracted set of high-level features,
instead of a cold start, being trained from a randomly initialized embedding space.
Based on such, a new representation can always be dynamically adjusted by the model for different downstream tasks.

\section{Conclusion}
The music generation task sees a large body of research recently,
utilizing different tweaked input encodings and diverse feature engineering techniques with improving results.
We are motivated by the monolithic model size, the inconsistent and taken-for-granted encoding approaches as used in the literature.
We present a systematic comparison of the different encoding options and encoding hyper-parameters,
based on the experiment results on a small Transformer-XL network of only 0.5M parameters.
Results suggests that the current Transformer-based auto-regressive generation systems are quite sensitive to these hyper-parameters,
which closely interact with the model despite that they are not a part of the model architecture.
Problems such as over-fitting are still observed for the tiny network with only 0.5M parameters.
Results also demonstrate the advantages and drawbacks of different encoding options,
so we recommend that different encoding options should be carefully chosen for an auto-regressive music generation model.
The findings of our works can also contribute to the latest generation models that are not in an auto-regressive manner,
which means different encoding options for the same feature could be incorporated to improve the performance.

\section{References}

\bibliographystyle{IEEEtranN}
\bibliography{MIR}
\end{document}